\documentclass[a4paper,twocolumn,10pt]{article}

\PassOptionsToPackage{hyphens}{url}
\usepackage[breaklinks]{hyperref}
\newcommand*{\Murl}[1]{\href{#1}{#1}}

\usepackage{xcolor}

\usepackage{savesym}
\savesymbol{iint}
\savesymbol{iiint}
\savesymbol{iiiint}
\savesymbol{idotsint}
\usepackage{amsmath}
\restoresymbol{AMS}{iint}
\restoresymbol{AMS}{iiint}
\restoresymbol{AMS}{iiiint}
\restoresymbol{AMS}{idotsint}

\let\oldfloatpagefraction\floatpagefraction
\let\olddblfloatpagefraction\dblfloatpagefraction
\renewcommand\floatpagefraction{.9}
\renewcommand\dblfloatpagefraction{.9} %

\usepackage[natbib,cref,color]{nikis}

\usepackage{booktabs}

\usepackage{tikz}
\usepackage{pgfplots}
\usepgfplotslibrary{colorbrewer}

\pgfplotsset{
cycle list/Dark2,
cycle multiindex* list={
mark list*\nextlist
Dark2\nextlist
},
}

\pgfplotsset{
width=150mm,height=100mm,
major grid style={thin,dotted,color=black!50},
minor grid style={thin,dotted,color=black!50},
grid,
every axis/.append style={
line width=0.5pt,
tick style={
line cap=round,
thin,
major tick length=4pt,
minor tick length=2pt,
},
},
legend cell align=left,
legend pos=north west,
}

\pgfplotsset{%
squeezedPlot/.style={%
width=0.5\linewidth,
height=0.5\linewidth
},
appendLegend/.style={%
transpose legend=true,
legend style={font=\small}}
}

\newcommand*{\bucketMax}{\ensuremath{b_{\textup{max}}}}
\newcommand*{\iAvx}{\textsf{avx}}
\newcommand*{\iCht}{\textsf{cht}}
\newcommand*{\iChmap}{\textsf{single}}
\newcommand*{\iPlain}{\textsf{plain}}

\newcommand*{\iPlainMD}{\ensuremath{\iPlain^{\%}_\textup{50}}}

\newcommand*{\iPlainMI}{\ensuremath{\iPlain^{\%}_\textup{\PlusPlus}}}

\newcommand*{\iAvxD}{\ensuremath{\iAvx_\textup{50}}}
\newcommand*{\iChtD}{\ensuremath{\iCht_\textup{50}}}

\newcommand*{\iPlainD}{\ensuremath{\iPlain_\textup{50}}}

\newcommand*{\iAvxI}{\ensuremath{\iAvx_\textup{\PlusPlus}}}
\newcommand*{\iChtI}{\ensuremath{\iCht_\textup{\PlusPlus}}}
\newcommand*{\iChmapI}{\ensuremath{\iChmap_\textup{\PlusPlus}}}
\newcommand*{\iPlainI}{\ensuremath{\iPlain_\textup{\PlusPlus}}}

\newcommand*{\iCleary}{\textsf{cleary}}
\newcommand*{\iElias}{\textsf{elias}}
\newcommand*{\iLayered}{\textsf{layered}}

\newcommand*{\iClearyP}{\ensuremath{\iCleary_{\textup{P}}}}
\newcommand*{\iEliasP}{\ensuremath{\iElias_{\textup{P}}}}
\newcommand*{\iLayeredP}{\ensuremath{\iLayered_{\textup{P}}}}

\newcommand*{\iClearyS}{\ensuremath{\iCleary_{\textup{S}}}}
\newcommand*{\iEliasS}{\ensuremath{\iElias_{\textup{S}}}}
\newcommand*{\iLayeredS}{\ensuremath{\iLayered_{\textup{S}}}}

\newcommand*{\iGoogle}{\textsf{google}}
\newcommand*{\iRigtorp}{\textsf{rigtorp}}
\newcommand*{\iSpp}{\textsf{spp}}
\newcommand*{\iTsl}{\textsf{tsl}}
\newcommand*{\iStd}{\textsf{std}}

\newcommand*{\effXor}[1]{f^{\otimes}_{#1}}
\newcommand*{\effMul}[1]{f^{\times}_{#1}}

\newcommand*{\myblock}[1]{\noindent{\textbf{#1}.}}

\ifcsname acknowledgment\endcsname
\else
\newenvironment{acknowledgment}{\paragraph{Acknowledgments}}{}
\fi

\title{Separate Chaining Meets Compact Hashing}
\begin{document}

\ifcsname affiliate\endcsname
\affiliate{FUKU}{Department of Informatics, Kyushu University, Japan Society for Promotion of Science}
\author{Dominik K\"{o}ppl}{FUKU}[dominik.koeppl@inf.kyushu-u.ac.jp]
\else
\author{Dominik K\"{o}ppl}
\date{Department of Informatics, Kyushu University, Japan Society for Promotion of Science}
\fi

\maketitle

\begin{abstract}
While separate chaining is a common strategy for resolving collisions in a hash table taught in most textbooks,
compact hashing is a less common technique for saving space when hashing integers whose domain is relatively small with respect to the problem size.
It is widely believed that hash tables waste a considerable amount of memory, as they either leave allocated space untouched (open addressing) or store additional pointers (separate chaining).
For the former, Cleary introduced the compact hashing technique that stores only a part of a key to save space.
However, as can be seen by the line of research focusing on compact hash tables with open addressing, there is additional information, called displacement, required for restoring a key.
There are several representations of this displacement information with different space and time trade-offs.
In this article, we introduce a separate chaining hash table that applies the compact hashing technique without the need for the displacement information.
Practical evaluations reveal that insertions in this hash table are faster or use less space than all previously known compact hash tables on modern computer architectures when storing sufficiently large satellite data.
\end{abstract}

\section{Introduction}
A major layout decision for hash tables is how collisions are resolved.
A well-studied and easy-implementable layout is separate chaining, which
is also applied by the hash table \texttt{unordered\_map} of the \CPlusPlus{} standard library \texttt{libstdc\PlusPlus}~\cite[Sect.~22.1.2.1.2]{gnu18cppmanual}.
On the downside, it is often criticized for being bloated and slow%
\footnote{Cf. \Murl{http://www.idryman.org/blog/2017/05/03/writing-a-damn-fast-hash-table-with-tiny-memory-footprints/},\\
\Murl{https://probablydance.com/2017/02/26/i-wrote-the-fastest-hashtable/},\\
\Murl{https://attractivechaos.wordpress.com/2018/10/01/advanced-techniques-to-implement-fast-hash-tables},\\
\Murl{https://tessil.github.io/2016/08/29/benchmark-hopscotch-map.html}, to name a few.}.
In fact, almost all modern replacements feature open addressing layouts.
Their implementations are highlighted with detailed benchmarks putting separate chaining with \texttt{unordered\_map} as its major representation in the backlight of interest.
However, when considering compact hashing with satellite data, separate chaining becomes again a competitive approach,
on which we shed a light in this article.

\subsection{Related Work}
The hash table of \citet{askitis09integerhash} also resorts to separate chaining. 
Its buckets are represented as dynamic arrays. On inserting an element into one of these array buckets,
the size of the respective array increments by one (instead of, e.g., doubling its space).
The approach differs from ours in that these arrays store a list of (key,value)-pairs
while our buckets separate keys from values.

The scan of the buckets in a separate chaining hash table can be accelerated with SIMD (single instruction multiple data) instructions as shown by
\citet{ross07simdhash} who studied the application of SIMD instructions
for comparing multiple keys in parallel in a \emph{bucketized Cuckoo hash table}.

For reducing the memory requirement a of hash table, a \emph{sparse} hash table layout was introduced by members of Google\footnote{\url{https://github.com/sparsehash/sparsehash}\label{footSparseHashMap}}.
Sparse hash tables are a lightweight alternative to standard open addressing hash tables, which are represented as plain arrays.
Most of the sparse variants replace the plain array with a bit vector of the same length marking positions at which an element would be stored in the array.
The array is emulated by this bit vector and its partitioning into buckets, which are dynamically resizeable and store the actual data.

The notion of \emph{compact hashing} was coined by \citet{cleary84cht} who studied a hash table with bidirectional linear probing.
The idea of compact hashing is to use an injective function mapping keys to pairs of integers.
Using one integer, called \emph{remainder}, as a hash value, and the other, called \emph{quotient}, as the data stored in the hash table,
the hash table can restore a key by maintaining its quotient and an information to retain its corresponding remainder.
This information, called \emph{displacement}, is crucial as the bidirectional linear probing displaces elements on a hash collision from the position corresponding to its hash value, i.e., its remainder.
\citet{poyias15bonsai} gave different representations for the displacement in the case that the hash table applies linear probing.

In this paper, we show that it is not necessary to store additional data in case that we resort to separate chaining as collision resolution.
The main strength of our hash table is its memory-efficiency during the construction while being at least as fast as other compact hash tables.
Its main weakness is the slow lookup time for keys, as we do not strive for small bucket sizes.

\section{Separate Chaining with Compact Hashing}
Our hash table~$H$ has $|H|$ buckets, where $|H|$ is a power of two.
Let $h$ be the hash function of~$H$.
An element with key~$K$ is stored in the $(h(K) \bmod |H|)$-th bucket.
To look up an element with key~$K$, the $(h(K) \bmod |H|)$-th bucket is linearly scanned.

A common implementation represents a bucket with a linked list,
and tries to avoid collisions as they are a major cause for decelerating searches.
Here, the buckets are realized as dynamic arrays, similar to the array hash table of \citet{askitis09integerhash}.
We further drop the idea of avoiding collisions.
Instead, we want to maintain buckets of sufficiently large sizes to compensate the extra memory for maintaining (a) the pointers to the buckets and (b) their sizes.
To prevent a bucket from growing too large, we introduce a threshold~\bucketMax{} for the maximum size.
Choosing an adequate value for \bucketMax{} is important, as it affects the resizing and the search time of our hash table.

\myblock{Resize}
When we try to insert an element into a bucket of maximum size~\bucketMax{}, we create a new hash table with twice the number of buckets~$2|H|$
and move the elements from the old table to the new one, bucket by bucket.
After a bucket of the old table becomes empty, we can free up its memory.
This reduces the memory peak commonly seen in hash tables or dynamic vectors reserving one large array,
as these data structures need to reserve space for $3m$ elements when resizing from $m$ to $2m$.
This technique is also common for sparse hash tables.

\begin{figure*}
\centering{%
\includegraphics[width=0.8\linewidth]{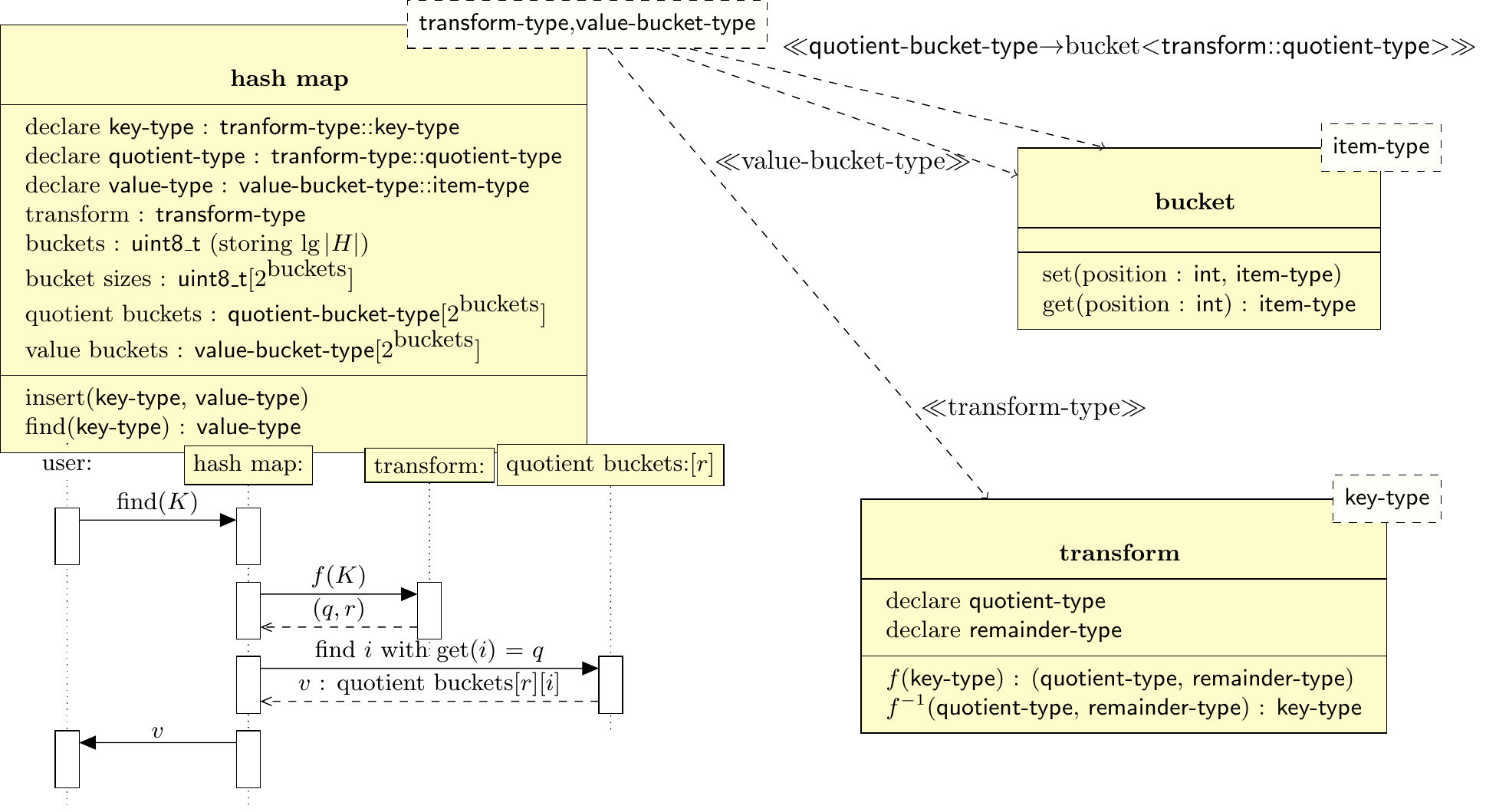}
}%
\caption{Diagram of our proposed hash table. 
The call of find($K$) returns the $i$-th element of the $r$-th bucket at which $K$ is located.
The injective transform determines the types of the key and the quotient.
}
\label{figUMLDiagram}
\end{figure*}

\myblock{Search in Cache Lines}
We can exploit modern computer architectures featuring large cache sizes
by selecting a sufficiently small \bucketMax{} such that buckets fit into a cache line.
Since we are only interested in the keys of a bucket during a lookup,
an optimization is to store keys and values separately:
In our hash table, a bucket is a composition of a key bucket and a value bucket, each of the same size.
This gives a good locality of reference~\cite{denning68locality} for searching a key.
This layout is favorable for large values of~\bucketMax{} and (keys,value)-pairs where the key size is relatively small to the value size,
since (a) the cost for an extra pointer to maintain two buckets instead of one becomes negligible while
(b) more keys fit into a cache line when searching a key in a bucket.
An overview of the proposed hash table layout is given in \cref{figUMLDiagram}.

\subsection{Compact Hashing}

Compact hashing restricts the choice of the hash function~$h$.
It requires an injective transform~$f$ that maps a key~$K$ to two integers~$(q,r)$ with
$1 \le r \le |H|$, where $r$ acts as the hash value~$h(K)$.
The values~$q$ and~$r$ are called \emph{quotient} and \emph{remainder}, respectively.
The quotient~$q$ can be used to restore the original key~$K$ if we know its corresponding remainder~$r$.
We translate this technique to our separate chaining layout by storing $q$ as key in the $r$-th bucket on inserting a key~$K$ with $f(K) = (q,r)$.

A discussion of different injective transforms is given by \citet[Sect.~3.2]{fischer17lz78}.
Suppose that all keys can be represented by $k$ bits.
We want to construct a bijective function $f : [1..2^k] \rightarrow f([1..2^k])$, where we use the last
$\lg m$ bits for the remainder and the other bits for the quotient.
Our used transform\footnote{\url{https://github.com/kampersanda/poplar-trie}} is inspired by the splitmix algorithm~\cite{steele14splitmix}.
It intermingles three xorshift~\cite{marsaglia03xorshift} functions~$\effXor{j} : x \mapsto x \mathop{\otimes} (2^j x) \mod 2^k$
with three multiplicative functions~$\effMul{c} : x \mapsto c x \mod 2^k$, where $\otimes$ denotes the bit-wise exclusive OR operation.
The composition of these functions is invertible, since each of them itself is invertible:

\myblock{Xorshift}
The function $\effXor{j}$ is self-inverse, i.e., $\effXor{j}\circ \effXor{j}(K) = K$, for an integer~$j$ with $k \ge j > \gauss{k/2}$ or $-k \le j < -\gauss{k/2}$.
For datasets whose keys only slightly differ (i.e., incremental values),
selecting $j < -\gauss{k/2}$ instead of $j > \gauss{k/2}$ is more advantageous since
the former distributes the last $k+j$ bits affecting the remainder.
This can lead to a more uniform distribution of the occupation of the buckets.

\myblock{Multiplicative}
Each of our functions~$\effMul{c}$ is initialized with an odd number $c$ less than $2^k$.
It is known that the family~$\menge{\effMul{c}}_{c~\text{odd}}$ is universal~\cite[Sect.~2.2]{dietzfelbinger97hash}, but not strongly universal~\cite{thorup00hash}.
Since $c$ and $2^k$ are relatively prime, there is an
modular multiplicative inverse of~$c$ with respect to the divisor~$2^k$, which
we can find with the extended Euclidean algorithm in \Oh{k} time in a precomputation step~\cite[Sect.~4.5.2]{knuthArt2Seminumerical}.

\subsection{Resize Policies}
We resize a bucket with the C function \texttt{realloc}.
Whether we need to resize a bucket on inserting an element depends on the policy we follow:

\begin{description}
\crampedItems{}
\item[Incremental Policy]: Increment the size of the bucket such that the new element just fits in.
This policy saves memory as only the minimum required amount of memory is allocated.
As buckets store at most $\bucketMax{} = \Oh{1}$ elements, the resize can be performed in constant time.
In practice, however, much of the spent time depends on the used memory allocator for increasing the allocated space.
We append \bsq{\PlusPlus} to a hash table in subscript if it applies this policy.
\item[Half Increase]: Increase the size of a bucket by 50\%\footnote{Inspired by the discussion in \url{https://github.com/facebook/folly/blob/master/folly/docs/FBVector.md}.}.
This policy eases the burden of the allocator at the expense of possibly wasting memory for unoccupied space in the buckets.
We append \bsq{50} in subscript to a hash table if it applies this policy.
\end{description}

\subsection{Bucket Variations}
Our hash table layout in \cref{figUMLDiagram} supports different quotient and value bucket types.
In the experiments, we call a hash table by the name of its quotient bucket representation.
There, we evaluated the following representations:

\myblock{\iCht{}}
Our default quotient bucket stores quotients bit-compactly,
i.e., it stores a quotient in $k - \lg |H|$ bits%
\footnote{More precisely, the quotient needs $\upgauss{\lg \abs{f}} - \lg \abs{H}$ bits, 
  where $\upgauss{\lg \abs{f}}$ is the number of bits needed to represent all values of the transform~$f$,
which is $k$ in our case.},
where $k$ is the bit size needed to represent all keys and $|H|$ is the number of buckets of~$H$.
For that, it uses bit operations to store the quotients in a byte array.
The number of bits used by a key bucket is quantized at eight bits (the last byte of the array might not be full).
Since $\lg |H|$ is constant until a rehashing occurs, we do not have to maintain this value in each bucket.

\myblock{\iChmap{}}
A variant storing keys and values in a single bucket instead of two separate ones
can save additional space.
However, this space improvement is marginal compared to the more severe slowdown
for either (a) locating a quotient if we maintain a list of (key,value)-pairs or
(b) changing the size of the bucket if we first store all keys and then subsequently all values.
For the experiments, we used the representation~(b).

\myblock{\iAvx{}}
Another representation of the quotient bucket applies SIMD instructions to speed up the search of a key in a large bucket.
For that, it restricts the quotients to be quantized at 8 bits.
We use the AVX2 instructions
\texttt{\_mm256\_set1\_epi$k$} and \texttt{\_mm256\_cmpeq\_epi$k$} for loading a quotient with $k$ bits and comparing this loaded value with the entries of the bucket, respectively.
The \texttt{realloc} function for resizing a bucket cannot be used in conjunction with \iAvx{}, 
since the allocated memory for \iAvx{} must be 32-byte aligned.

\myblock{\iPlain{}}
For comparison, we also implemented a variant that does not apply compact hashing.
For that, we created the trivial injective transform $f_h : K \mapsto (K, h(K) \bmod 2^m)$
that uses an arbitrary hash function~$h$ for computing the remainders while producing quotients equal to the original keys.
Its inverse is trivially given by $f^{-1}(q,r) = q$.

\begin{figure*}
\pgfplotsset{squeezedPlot}

\begin{tikzpicture}
\begin{axis}[
title={Construction Time},
xlabel={number of elements [$\lg$]},
ylabel={avg.\ time per element [$\mu$s/\#]},
appendLegend,
legend to name={legRandomInsertTime},
legend columns=3
]

\addplot coordinates { (12.3399,10.7095) (12.9248,13.3763) (13.5098,11.5678) (14.0947,13.544) (14.6797,12.3212) (15.2647,15.5522) (15.8496,19.9206) (16.4346,19.8074) };
\addlegendentry{\iAvxD};
\addplot coordinates { (12.3399,61.6271) (12.9248,89.3297) (13.5098,72.3122) (14.0947,85.7539) (14.6797,66.0738) (15.2647,59.7427) (15.8496,66.7378) (16.4346,56.9215) };
\addlegendentry{\iAvxI};
\addplot coordinates { (12.3399,40.7155) (12.9248,60.5648) (13.5098,57.8331) (14.0947,68.0121) (14.6797,57.6276) (15.2647,55.5345) (15.8496,56.1092) (16.4346,50.6662) };
\addlegendentry{\iChtI};
\addplot coordinates { (12.3399,16.8997) (12.9248,13.5004) (13.5098,15.7362) (14.0947,13.0988) (14.6797,15.0859) (15.2647,13.7633) (15.8496,18.8953) (16.4346,25.2768) };
\addlegendentry{\iClearyP};
\addplot coordinates { (12.3399,44.7867) (12.9248,54.4033) (13.5098,45.541) (14.0947,43.604) (14.6797,41.6632) (15.2647,28.0542) (15.8496,36.373) (16.4346,33.018) };
\addlegendentry{\iGoogle};
\addplot coordinates { (12.3399,45.5615) (12.9248,36.8259) (13.5098,44.313) (14.0947,37.2303) (14.6797,48.0773) (15.2647,40.4395) (15.8496,51.2048) (16.4346,58.5063) };
\addlegendentry{\iLayeredS};
\addplot coordinates { (12.3399,13.1223) (12.9248,16.3858) (13.5098,14.7254) (14.0947,17.1966) (14.6797,15.743) (15.2647,19.6543) (15.8496,23.9941) (16.4346,22.7726) };
\addlegendentry{\iPlainD};
\addplot coordinates { (12.3399,6.01194) (12.9248,5.3961) (13.5098,5.78076) (14.0947,6.56099) (14.6797,8.40889) (15.2647,12.0826) (15.8496,12.9844) (16.4346,17.6153) };
\addlegendentry{\iRigtorp};
\addplot coordinates { (12.3399,14.7382) (12.9248,13.7846) (13.5098,13.8897) (14.0947,16.0882) (14.6797,16.5122) (15.2647,17.8647) (15.8496,20.3482) (16.4346,22.9145) };
\addlegendentry{\iSpp};

\end{axis}
\end{tikzpicture}
\begin{tikzpicture}
\begin{axis}[
title={Query Time},
xlabel={number of elements [$\lg$]},
ylabel={avg.\ time per element [$\mu$s/\#]}
]

\addplot coordinates { (12.3399,5.08578) (12.9248,5.07243) (13.5098,5.35628) (14.0947,5.26171) (14.6797,5.6696) (15.2647,5.78172) (15.8496,6.79231) (16.4346,9.55205) };
\addlegendentry{\iAvxD};
\addplot coordinates { (12.3399,5.09458) (12.9248,5.03796) (13.5098,5.36373) (14.0947,5.1715) (14.6797,5.63451) (15.2647,5.79909) (15.8496,6.69061) (16.4346,9.4519) };
\addlegendentry{\iAvxI};
\addplot coordinates { (12.3399,12.622) (12.9248,10.4298) (13.5098,13.5676) (14.0947,11.3116) (14.6797,14.777) (15.2647,12.592) (15.8496,11.3654) (16.4346,16.1558) };
\addlegendentry{\iChtI};
\addplot coordinates { (12.3399,8.46304) (12.9248,8.17659) (13.5098,8.79621) (14.0947,8.14981) (14.6797,8.73525) (15.2647,9.4543) (15.8496,11.7207) (16.4346,14.294) };
\addlegendentry{\iClearyP};
\addplot coordinates { (12.3399,7.61545) (12.9248,6.43759) (13.5098,7.0335) (14.0947,6.71536) (14.6797,6.53799) (15.2647,7.45872) (15.8496,8.33538) (16.4346,11.2518) };
\addlegendentry{\iGoogle};
\addplot coordinates { (12.3399,5.36061) (12.9248,5.7201) (13.5098,5.73162) (14.0947,5.57205) (14.6797,5.79988) (15.2647,6.41002) (15.8496,7.93001) (16.4346,10.1169) };
\addlegendentry{\iLayeredS};
\addplot coordinates { (12.3399,7.63937) (12.9248,7.16391) (13.5098,7.96337) (14.0947,7.52698) (14.6797,8.62765) (15.2647,8.17012) (15.8496,8.92369) (16.4346,11.9027) };
\addlegendentry{\iPlainD};
\addplot coordinates { (12.3399,3.11574) (12.9248,3.3921) (13.5098,3.3043) (14.0947,3.21133) (14.6797,3.44109) (15.2647,4.00917) (15.8496,5.80877) (16.4346,7.58309) };
\addlegendentry{\iRigtorp};
\addplot coordinates { (12.3399,3.38717) (12.9248,3.71499) (13.5098,3.73263) (14.0947,3.75847) (14.6797,4.17032) (15.2647,4.52967) (15.8496,6.02528) (16.4346,7.75298) };
\addlegendentry{\iSpp};

\legend{}
\end{axis}
\end{tikzpicture}

\begin{minipage}{0.6\linewidth}
\caption{%
\parbox{0.9\linewidth}{%
  \emph{Left}: Time for inserting $2^{10} \cdot (3/2)^n$ elements consisting of randomly generated 32-bit values and 32-bit keys into a hash table, for $n \ge 0$.
\emph{Right:} Time for querying all inserted elements.
}%
}
\label{figRandomInsertTime}
\end{minipage}
\hfill
\begin{minipage}{0.35\linewidth}
\hfill
\ref{legRandomInsertTime}
\end{minipage}
\end{figure*}
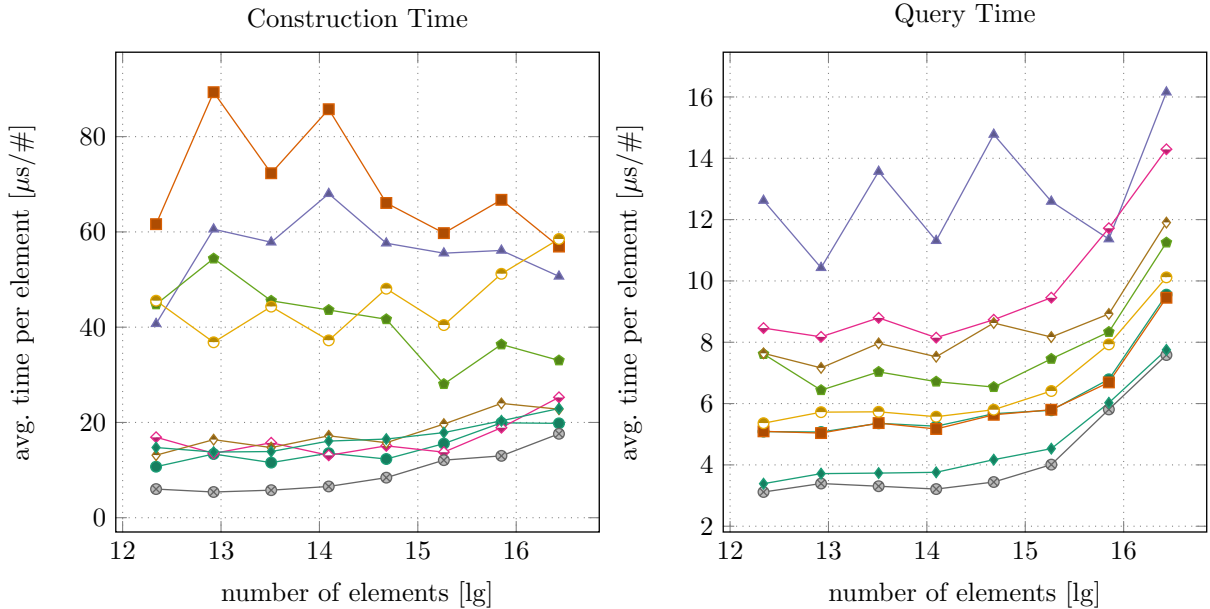

\subsection{Details on Sizes}

We set the maximum bucket size \bucketMax{} to 255 elements such that
we can represent the size of a bucket in a single byte.
A full bucket with 64-bit integers takes roughly 2KiB of memory, fitting in the L1 cache of a modern CPU.

For each bucket we store its size and pointers to its quotient and value bucket, using altogether 17 bytes.
Since we additionally waste less than one byte in \iCht{} for storing the quotients in a byte array, this gives an overhead of at most 18 bytes per bucket.
Let $m$ denote the (fractional) number of bytes needed to store an element.
Then our hash table uses $18 |H| + nm$ bytes for storing $n$ elements, 
where $|H|$ is at most $\upgauss{2n/\bucketMax{}}$ if we assume a uniform distribution of the elements among all buckets.

A non-sparse open addressing hash table with maximum load factor~$\alpha \le 1$
uses at least $nm/\alpha$ bytes.
If $m \ge 3.03$ bytes,
we need to set $\alpha$ to more than $0.956$ to make the open addressing hash table slimmer than our separate hashing table.
When resorting to linear probing we encounter
\begin{equation}\label{eqCollision}
  c_\alpha := (1/2) \cdot ( 1 + (1/(1-\alpha))^2)
\end{equation}
collisions on average for an insertion operation~\cite[Sect.~6.4]{knuthArt3Sorting}.
But $c_\alpha < \bucketMax{} \Leftrightarrow \alpha < 1 - 1/\sqrt{2\bucketMax - 1} \approx 0.956$, and hence 
such a table faces more collisions on average or uses more space than our proposed hash tables.
If $m < 3.03$ represents the number of bytes for storing a key and a value, 
one would usually resort to storing the data in a plain array, as there can be at most $3^{8m} \approx 20 \cdot 10^6$ keys.
The only interesting domain is when we consider compact hashing for $m < 3.03$,
where $m$ now represents the number of bytes we need for the \emph{quotient} and the value.
However, compact representations of open addressing hash tables need to store displacement information,
which should take at most $18 \abs{H}\text{~bytes} \le 1.13\text{~bits}$ per element to be on par with the memory overhead of the separate chaining layout.

\section{Experiments}
We implemented our proposed hash table in \CPlusPlus{}17.
The implementation is freely available at \url{https://github.com/koeppl/separate_chaining}.

\myblock{Evaluation Setting}
Our experiments ran on an Ubuntu Linux 18.04 machine equipped with 32 GiB of RAM and an Intel Xeon CPU E3-1271 v3 clocked at 3.60GHz.
We measure the memory consumption by overloading the calls to  \texttt{malloc}, \texttt{realloc}, \texttt{free} and its \CPlusPlus{} counterparts with the tudostats framework\footnote{\url{https://github.com/tudocomp/tudostats}}.
The benchmark was compiled with the flags \texttt{-O3 -DNDEBUG --march=native}, the last option for supporting AVX2 instructions.

\myblock{Contestants}
We selected the following hash tables that are representative \CPlusPlus{} hash tables, are sparse, or work with compact hashing.
\begin{itemize}
\crampedItems{}
\item \iStd{}: The \texttt{unordered\_map} implementation of \texttt{libstdc\PlusPlus}.
We used the default maximum load factor $1.0$, i.e., we resize the hash table after the number of stored elements exceeds the number of buckets.
\item \iRigtorp{}: The fast but memory-hungry linear-probing hash table of Erik Rigtorp\footnote{\url{https://github.com/rigtorp/HashMap}}.
The load factor is hard-coded to $0.5$.
\item \iGoogle{}: Google's sparse hash table$^{{\ref{footSparseHashMap}}}$ with quadratic probing. Its maximum load factor is set to the default value~$0.8$.
\item \iSpp{}: Gregory Popovitch's Sparsepp\footnote{\url{https://github.com/greg7mdp/sparsepp}}, a derivate of Google's sparse hash table. Its maximum load factor is $0.5$.
\item \iTsl{}: Tessil's sparse map\footnote{\url{https://github.com/Tessil/sparse-map}} with quadratic probing. Its default maximum load factor is $0.5$.
\item \iCleary{}, \iElias{}, \iLayered{}: The compact hash tables of \citet{cleary84cht} and \citet{poyias15bonsai}.
  \begin{itemize}
    \item \iElias{} partitions the displacement into integer arrays of length 1024, which are encoded with Elias-$\gamma$~\cite{elias74code}.
    \item \iLayered{} stores this information in two multiple associative array data structures.
      The first is an array storing 4-bit integers, and the second is an \texttt{unordered\_map} for displacements larger than 4 bits.
  \end{itemize}
The implementations are provided by the tudocomp project\footnote{\url{https://github.com/tudocomp/compact_sparse_hash}}.
All hash tables apply linear probing, and support a sparse table layout.
We call these hash tables \emph{Bonsai tables} for the following evaluation,
and append in subscript \bsq{P} or \bsq{S} if the respective variant is in its plain form or in its sparse form, respectively.
We used the default maximum load factor of $0.5$.
\end{itemize}

\begin{figure*}
\pgfplotsset{squeezedPlot}

\begin{tikzpicture}
\begin{axis}[
title={Random Insertion : Space},
xlabel={number of elements [$\lg$]},
ylabel={avg.\ memory per element [bytes/\#]}
]

\addplot coordinates { (24.0391,5.94677) (24.624,5.92387) (25.209,5.81337) (25.794,5.71005) (26.3789,5.6817) (26.9639,5.5741) (27.5489,5.5494) };
\addlegendentry{\iChmap};
\addplot coordinates { (24.0391,6.0076) (24.624,6.04553) (25.209,5.86744) (25.794,5.78215) (26.3789,5.72976) (26.9639,5.63818) (27.5489,5.59212) };
\addlegendentry{\iChtI};
\addplot coordinates { (24.0391,7.54294) (24.624,6.21772) (25.209,6.58372) (25.794,5.92562) (26.3789,6.21805) (26.9639,5.64437) (27.5489,5.91837) };
\addlegendentry{\iClearyS};
\addplot coordinates { (24.0391,7.03411) (24.624,5.99837) (25.209,6.23208) (25.794,5.74487) (26.3789,5.93483) (26.9639,5.49952) (27.5489,5.67608) };
\addlegendentry{\iEliasS};
\addplot coordinates { (24.0391,8.64885) (24.624,8.43257) (25.209,8.57676) (25.794,8.76901) (26.3789,8.51267) (26.9639,8.68356) (27.5489,8.45571) };
\addlegendentry{\iGoogle};
\addplot coordinates { (24.0391,9.01965) (24.624,6.86773) (25.209,7.89611) (25.794,6.50557) (26.3789,6.98717) (26.9639,6.16744) (27.5489,6.60229) };
\addlegendentry{\iLayeredS};
\addplot coordinates { (24.0391,9.67448) (24.624,10.5593) (25.209,10.5184) (25.794,10.4854) (26.3789,10.4867) (26.9639,9.71318) (27.5489,9.44704) };
\addlegendentry{\iPlainD};
\addplot coordinates { (24.0391,13.0928) (24.624,10.3101) (25.209,11.6379) (25.794,10.0325) (26.3789,10.819) (26.9639,9.79814) (27.5489,10.4477) };
\addlegendentry{\iSpp};
\addplot coordinates { (24.0391,11.435) (24.624,9.78401) (25.209,10.3789) (25.794,9.58592) (26.3789,10.1144) (26.9639,9.40984) (27.5489,9.87978) };
\addlegendentry{\iTsl};

\end{axis}
\end{tikzpicture}
\begin{tikzpicture}

\begin{axis}[
title={Random Insertion : Time/Space},
xlabel={avg.\ time per element [s/\#]},
ylabel={avg.\ memory per element [bytes/\#]},
legend pos=north east,
legend columns=3
]

\addplot coordinates { (0.0964657,9.90625) (0.09822,9.74132) (0.145695,9.80729) (0.177203,8.18142) (0.232746,8.69273) (0.24751,7.90177) (0.255767,8.76517) (0.267993,8.6548) (0.271993,8.69433) (0.279178,7.65727) (0.279688,9.24595) (0.302497,8.05601) (0.311917,8.51781) (0.31914,8.5374) (0.320552,7.56949) (0.336352,9.8451) (0.348158,8.68731) (0.350189,7.71781) (0.352405,7.71327) (0.358203,8.46931) (0.377872,8.15481) (0.409172,8.07944) (0.422337,7.47996) (0.444129,7.52527) (0.476806,7.04129) (0.485912,7.50731) (0.510692,7.50952) (0.525318,7.32601) (0.534991,7.30462) (0.581444,6.46218) (0.583578,6.63532) };
\addlegendentry{\iChtD};
\addplot coordinates { (0.148788,7.62717) (0.164407,7.78516) (0.241263,7.6901) (0.253692,7.58478) (0.333501,7.34688) (0.345647,7.21045) (0.37172,7.39545) (0.382564,6.97599) (0.3911,7.25389) (0.394129,6.83928) (0.396401,7.11096) (0.436233,6.88399) (0.437539,6.60505) (0.445981,7.02633) (0.451407,6.74456) (0.456577,6.46941) (0.483496,7.48592) (0.50197,6.51662) (0.508486,6.65758) (0.517275,6.37545) (0.576531,6.23689) (0.62356,6.29281) (0.695769,6.14542) (0.701701,6.09695) (0.759549,6.0076) (0.795264,5.86744) (0.837421,6.04553) (0.842745,5.72976) (0.8955,5.78215) (0.898112,5.59212) (0.939677,5.63818) };
\addlegendentry{\iChtI};
\addplot coordinates { (0.376963,7.13035) (0.377186,7.4393) (0.394988,7.77083) (0.411812,7.71811) (0.41403,7.37235) (0.43633,7.09674) (0.437218,8.04321) (0.44754,7.7537) (0.451727,7.4065) (0.458073,6.80495) (0.464435,6.54598) (0.482048,8.41289) (0.496332,8.08854) (0.504189,8.61458) (0.509998,7.15319) (0.519953,7.46182) (0.526371,8.50887) (0.528671,6.82729) (0.577903,6.51778) (0.615104,6.2148) (0.630727,7.49554) (0.644864,7.9375) (0.656674,6.82192) (0.709485,6.21772) (0.711089,6.52356) (0.724799,5.92562) (0.756299,5.64437) (0.796864,6.21805) (0.804341,5.91837) (0.808696,6.58372) (0.809318,7.54294) };
\addlegendentry{\iClearyS};
\addplot coordinates { (0.179845,8.46776) (0.186228,8.52469) (0.189691,8.49315) (0.191633,8.55479) (0.197415,8.59028) (0.199366,8.62369) (0.212614,8.70165) (0.21628,8.4375) (0.218432,8.73972) (0.22399,8.43844) (0.22632,8.87963) (0.229497,9.25408) (0.238061,8.65766) (0.238527,8.65625) (0.252131,8.58459) (0.263427,8.51966) (0.295384,8.77949) (0.306079,8.46192) (0.315328,8.69288) (0.357459,8.6159) (0.40711,8.54746) (0.422453,9.11544) (0.456962,8.48664) (0.478838,8.72996) (0.506919,8.43257) (0.530122,8.64885) (0.558483,8.57676) (0.569926,8.51267) (0.587779,8.45571) (0.594061,8.76901) (0.620818,8.68356) };
\addlegendentry{\iGoogle};
\addplot coordinates { (0.354867,7.98868) (0.358435,7.69368) (0.363753,8.37731) (0.371616,7.99939) (0.391274,8.42044) (0.417027,7.75533) (0.421453,8.83333) (0.424913,8.65986) (0.431294,8.14641) (0.439861,7.39409) (0.445001,9.97569) (0.453638,7.07555) (0.461727,9.84042) (0.46345,8.75521) (0.496792,7.93292) (0.497694,8.76119) (0.497918,10.0074) (0.516775,7.52071) (0.57402,7.1359) (0.605663,8.58594) (0.614272,6.76813) (0.618959,8.89609) (0.646658,7.79059) (0.705775,6.86773) (0.718949,6.50557) (0.738537,6.16744) (0.747124,7.25373) (0.78163,6.60229) (0.786096,6.98717) (0.789747,7.89611) (0.79749,9.01965) };
\addlegendentry{\iLayeredS};
\addplot coordinates { (0.0718662,9.859375) (0.0988511,10.7569) (0.106472,9.40104) (0.119022,10.6698) (0.120041,9.68038) (0.126492,8.9838) (0.13158,9.91038) (0.132922,10.1681) (0.136791,10.6378) (0.142948,10.0535) (0.151133,10.2686) (0.152795,9.18927) (0.153461,10.5649) (0.160867,9.72158) (0.183818,10.3993) (0.197276,10.5832) (0.219819,9.56995) (0.220691,10.5421) (0.227754,9.80674) (0.271118,9.95291) (0.326131,10.663) (0.343388,10.7122) (0.387335,10.0399) (0.391518,10.1341) (0.45901,9.67448) (0.482898,10.5184) (0.521835,10.4854) (0.523374,10.5593) (0.526188,10.4867) (0.576525,9.71318) (0.58802,9.44704) };
\addlegendentry{\iPlainD};
\addplot coordinates { (0.144959,8.28125) (0.188091,8.24306) (0.208771,8.1875) (0.218305,8.18793) (0.234014,8.16204) (0.236618,8.19672) (0.242732,8.21296) (0.245784,8.22151) (0.246817,8.24966) (0.259949,8.28189) (0.262367,8.17478) (0.28271,8.29508) (0.290015,8.33227) (0.290344,8.26217) (0.31813,8.23299) (0.322704,8.20708) (0.370356,8.31062) (0.370697,8.18406) (0.387476,8.27609) (0.459286,8.2454) (0.564865,8.21813) (0.573898,8.3272) (0.677353,8.1939) (0.677625,8.29084) (0.773116,8.25853) (0.820175,8.2298) (0.869522,8.20427) (0.885922,8.40682) (0.902392,8.3064) (0.930441,8.18157) (0.942938,8.27236) };
\addlegendentry{\iPlainI};
\addplot coordinates { (0.0888579,9.85082) (0.0896019,10.0486) (0.0949922,10.2067) (0.0959961,10.5453) (0.10441,9.82031) (0.106552,9.95228) (0.107959,10.9028) (0.109446,11.2101) (0.112529,11.9792) (0.116927,12.6127) (0.130966,10.6885) (0.13284,10.401) (0.150466,10.3481) (0.168207,13.2699) (0.171991,10.0619) (0.179909,11.8002) (0.193135,10.8681) (0.200317,9.82376) (0.231851,10.4905) (0.311899,10.1826) (0.31762,12.4281) (0.351305,9.92443) (0.362654,11.0613) (0.391,10.6474) (0.426677,10.3101) (0.429619,10.0325) (0.446903,13.0928) (0.45519,9.79814) (0.466369,11.6379) (0.474512,10.819) (0.496741,10.4477) };
\addlegendentry{\iSpp};
\addplot coordinates { (0.0630927,9.625) (0.0631068,9.66667) (0.0644861,9.48148) (0.0711327,9.95885) (0.0712847,10.6944) (0.0720991,9.72169) (0.0725579,10.1543) (0.0726771,9.52429) (0.0766919,9.85417) (0.0816624,10.3074) (0.0826626,11.0361) (0.0890809,10.035) (0.123971,9.81526) (0.132883,11.5982) (0.14577,9.60825) (0.14687,10.4119) (0.153766,10.1465) (0.155893,9.42941) (0.17312,9.90746) (0.198922,9.69424) (0.2,10.8574) (0.220401,10.2591) (0.23216,9.50436) (0.248528,10.0064) (0.269153,9.78401) (0.289096,9.58592) (0.292227,11.435) (0.30702,10.3789) (0.315899,10.1144) (0.318616,9.40984) (0.326011,9.87978) };
\addlegendentry{\iTsl};

\end{axis}
\end{tikzpicture}

\caption{\emph{Left}: Space needed for constructing the hash tables in the setting of \cref{figRandomInsertTime}.
\emph{Right}: Memory and time divided by the number of stored elements. Each element is composed of a 32-bit key and a 32-bit value, using combined 8 bytes.}
\label{figRandomInsertConstruction}
\end{figure*}
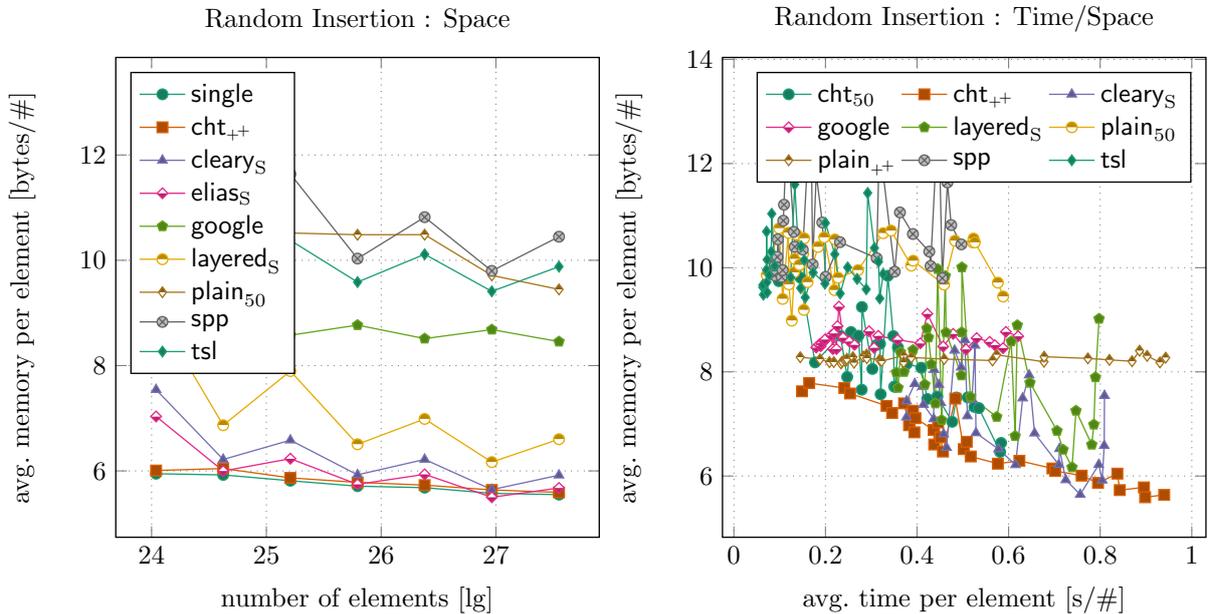

\subsection{Micro-Benchmarks}
Our micro benchmarks are publicly available at \url{https://github.com/koeppl/hashbench} for third-party evaluations.
We provide benchmarks for insertions, deletions, and lookups of (a) inserted keys (for successful searches) and (b) keys that are not present in the hash tables (for unsuccessful searches).

\myblock{Inserting Random Elements}
We used \texttt{std::rand} as a random generator to produce 32-bit keys and values.
The measured times and memory consumptions are depicted in \cref{figRandomInsertTime} and \cref{figRandomInsertConstruction}.
Our variants \iPlain{} and \iAvx{} do not use the compact hashing technique.
Instead, like other non-compact hash tables, they use the identity function in this setting.
Here, \iAvxD{} is faster than \iPlainD{} during the construction, 
and far superior when it comes to searching keys.

While the discrepancy in time between the incremental and the half increase policy is small for most bucket representations,
the construction of \iAvxI{} takes considerably longer than \iAvxD{}, as we cannot resort to the fast \texttt{realloc} for allocating \emph{aligned}
memory required for the SIMD operations.

The construction of \iChmap{} is tedious, as it needs to move all values of a bucket on each insertion.
On the other hand, its search time is on par with \iChtI{}.
Our compact and non-compact hash tables match the speed of the sparse and non-sparse Bonsai tables, respectively.

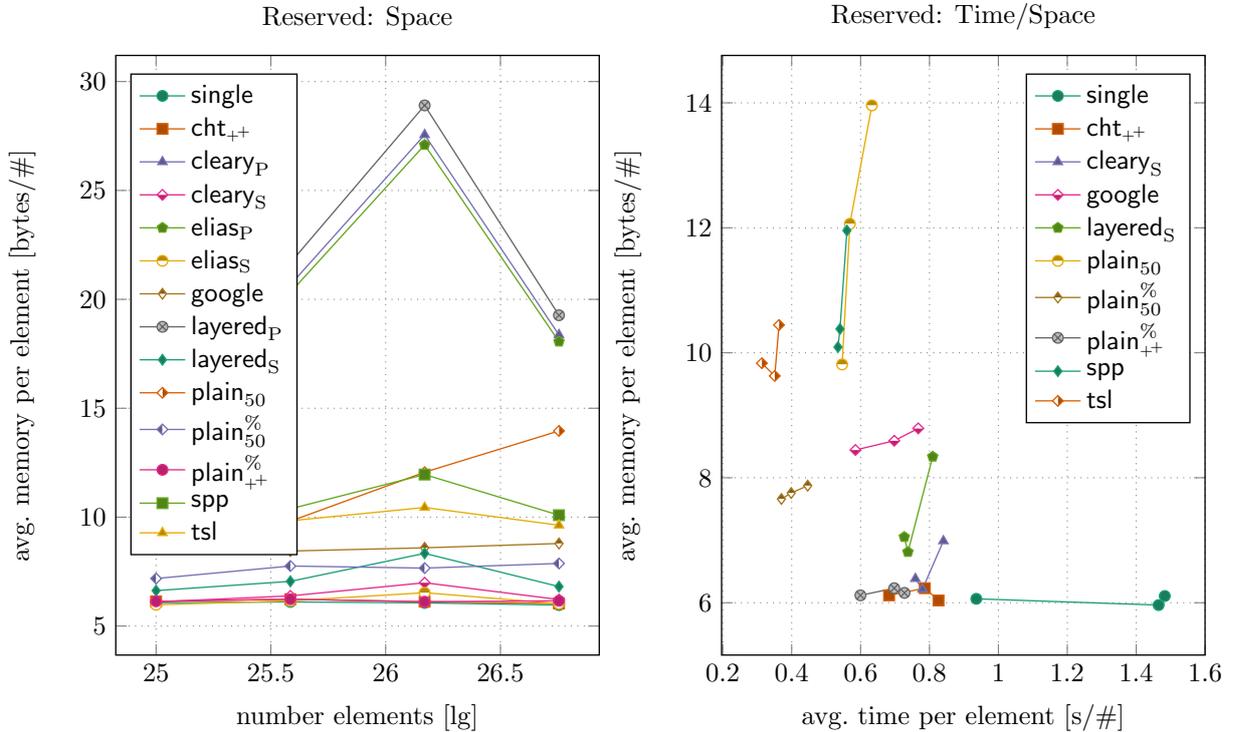
\begin{figure*}

\pgfplotsset{%
width=0.5\linewidth,
height=0.6\linewidth
}

\begin{tikzpicture}
\begin{axis}[
title={Reserved: Space},
xlabel={number elements [$\lg$]},
ylabel={avg.\ memory per element [bytes/\#]},
legend pos=north west,
]
\addplot coordinates { (25.0,6.07031) (25.585,6.1074) (26.1699,6.0625) (26.7549,5.96238) };
\addlegendentry{\iChmap};
\addplot coordinates { (25.0,6.13281) (25.585,6.2324) (26.1699,6.11806) (26.7549,6.03646) };
\addlegendentry{\iChtI};
\addplot coordinates { (25.0,15.875) (25.585,20.6667) (26.1699,27.5556) (26.7549,18.3704) };
\addlegendentry{\iClearyP};
\addplot coordinates { (25.0,6.10934) (25.585,6.3846) (26.1699,6.98823) (26.7549,6.21698) };
\addlegendentry{\iClearyS};
\addplot coordinates { (25.0,15.6142) (25.585,20.3185) (26.1699,27.0902) (26.7549,18.0601) };
\addlegendentry{\iEliasP};
\addplot coordinates { (25.0,5.97611) (25.585,6.15639) (26.1699,6.53095) (26.7549,6.02738) };
\addlegendentry{\iEliasS};
\addplot coordinates { (25.0,8.66667) (25.585,8.44444) (26.1699,8.59259) (26.7549,8.79012) };
\addlegendentry{\iGoogle};
\addplot coordinates { (25.0,16.6338) (25.585,21.6783) (26.1699,28.9044) (26.7549,19.2696) };
\addlegendentry{\iLayeredP};
\addplot coordinates { (25.0,6.6267) (25.585,7.05208) (26.1699,8.33704) (26.7549,6.81224) };
\addlegendentry{\iLayeredS};
\addplot coordinates { (25.0,10.4817) (25.585,9.8128) (26.1699,12.0642) (26.7549,13.9587) };
\addlegendentry{\iPlainD};
\addplot coordinates { (25.0,7.18205) (25.585,7.75572) (26.1699,7.66042) (26.7549,7.87352) };
\addlegendentry{\iPlainMD};
\addplot coordinates { (25.0,6.13281) (25.585,6.23241) (26.1699,6.11806) (26.7549,6.15741) };
\addlegendentry{\iPlainMI};
\addplot coordinates { (25.0,9.7531) (25.585,10.3812) (26.1699,11.9559) (26.7549,10.0912) };
\addlegendentry{\iSpp};
\addplot coordinates { (25.0,9.37508) (25.585,9.83318) (26.1699,10.4446) (26.7549,9.62932) };
\addlegendentry{\iTsl};

\end{axis}
\end{tikzpicture}
\begin{tikzpicture}
\begin{axis}[
title={Reserved: Time/Space},
xlabel={avg.\ time per element [s/\#]},
ylabel={avg.\ memory per element [bytes/\#]},
legend pos=north east,
]
\addplot coordinates { (0.935927,6.0625) (1.46507,5.96238) (1.483,6.1074) };
\addlegendentry{\iChmap};
\addplot coordinates { (0.682741,6.11806) (0.786388,6.2324) (0.826187,6.03646) };
\addlegendentry{\iChtI};
\addplot coordinates { (0.759038,6.3846) (0.779576,6.21698) (0.840352,6.98823) };
\addlegendentry{\iClearyS};
\addplot coordinates { (0.585072,8.44444) (0.697816,8.59259) (0.767388,8.79012) };
\addlegendentry{\iGoogle};
\addplot coordinates { (0.726862,7.05208) (0.737416,6.81224) (0.809312,8.33704) };
\addlegendentry{\iLayeredS};
\addplot coordinates { (0.546449,9.8128) (0.568968,12.0642) (0.63285,13.9587) };
\addlegendentry{\iPlainD};
\addplot coordinates { (0.37013,7.66042) (0.398647,7.75572) (0.446503,7.87352) };
\addlegendentry{\iPlainMD};
\addplot coordinates { (0.599666,6.11806) (0.697707,6.23241) (0.727423,6.15741) };
\addlegendentry{\iPlainMI};
\addplot coordinates { (0.534344,10.0912) (0.540166,10.3812) (0.560407,11.9559) };
\addlegendentry{\iSpp};
\addplot coordinates { (0.313507,9.83318) (0.350857,9.62932) (0.36322,10.4446) };
\addlegendentry{\iTsl};

\end{axis}
\end{tikzpicture}

\caption{%
Time for inserting $2^{25} \cdot (3/2)^n$ random elements for integers~$n \ge 0$.
The hash tables are prepared to reserve $2^{9} \cdot (3/2)^n$ buckets before the insertions start. }
\label{figMemoryReserve}
\end{figure*}

\myblock{Reversed Space}
Like in the previous experiment, we fill the hash tables with $n$ random elements for $n \ge 2^{16}$.
However, this time we let the hash tables reserve $2^{16}$ buckets in advance.
We added a percent sign in superscript to the \iPlain{} hash tables that (a) use our injective transform and (b) take (additionally) advantage of the fact that they only need to store quotients of at most 16 bits.
The results are visualized in \cref{figMemoryReserve}.
We see that \iPlainMD{} is superior to \iGoogle{}, while \iPlainMI{} uses far less space that other non-compact hash tables.
Like in \cref{figRandomInsertTime}, a major boost for lookups can be perceived if we exchange \iPlain{} with \iAvx{}%
\footnote{\iAvx{} is not shown in \cref{figMemoryReserve} since our memory allocation counting library does not count aligned allocations needed for \iAvx{}.}, which takes the same amount of space as \iPlain{}.

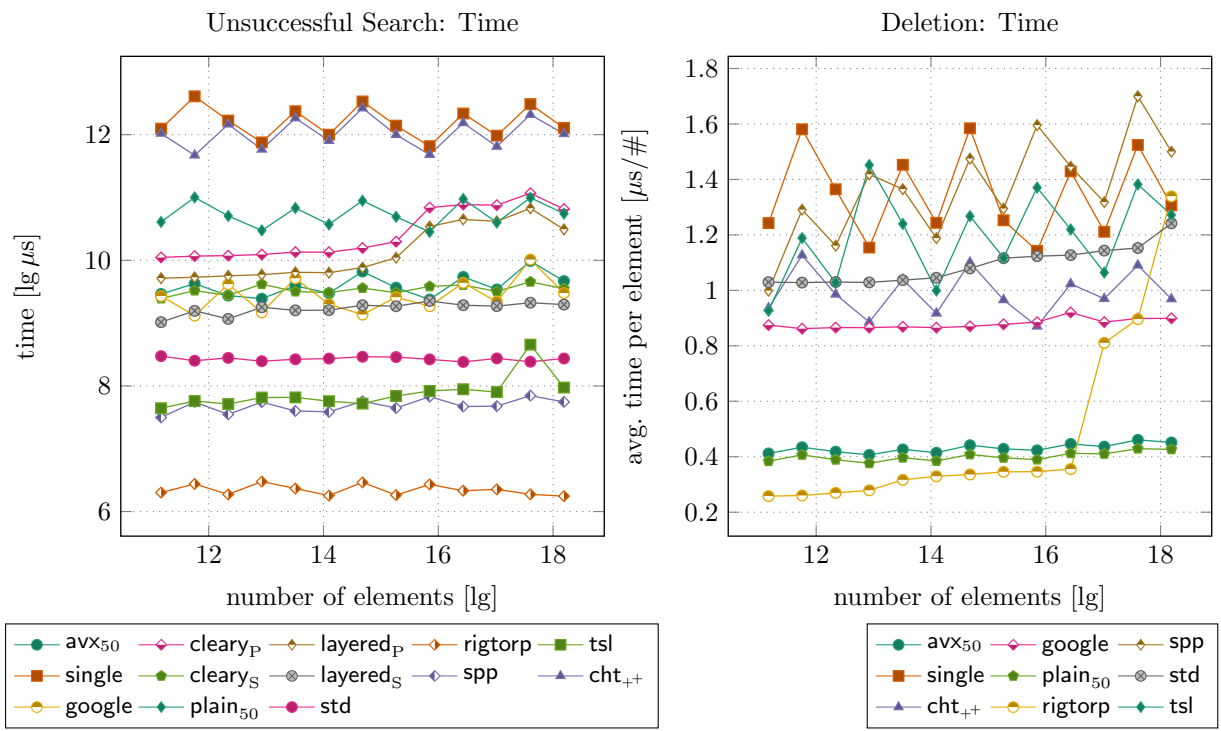
\begin{figure*}
\pgfplotsset{squeezedPlot}

\begin{tikzpicture}
\begin{axis}[
title={Unsuccessful Search: Time},
xlabel={number of elements [$\lg$]},
ylabel={time [$\lg \mu$s]},
appendLegend,
legend to name={legMissTime},
legend columns=3
]

\addplot coordinates { (11.1699,9.46387) (11.7549,9.62796) (12.3399,9.43748) (12.9248,9.38931) (13.5098,9.59364) (14.0947,9.47553) (14.6797,9.82034) (15.2647,9.56409) (15.8496,9.37584) (16.4346,9.73561) (17.0195,9.5367) (17.6045,9.98822) (18.1895,9.66912) };
\addlegendentry{\iAvxD};
\addplot coordinates { (11.1699,12.0956) (11.7549,12.6104) (12.3399,12.2253) (12.9248,11.8769) (13.5098,12.3735) (14.0947,11.9996) (14.6797,12.5285) (15.2647,12.1433) (15.8496,11.8169) (16.4346,12.3371) (17.0195,11.9836) (17.6045,12.4859) (18.1895,12.1077) };
\addlegendentry{\iChmap};
\addplot coordinates { (11.1699,12.0193) (11.7549,11.672) (12.3399,12.1657) (12.9248,11.7659) (13.5098,12.2635) (14.0947,11.9041) (14.6797,12.4223) (15.2647,11.9988) (15.8496,11.6817) (16.4346,12.1874) (17.0195,11.8121) (17.6045,12.3193) (18.1895,12.0131) };
\addlegendentry{\iChtI};
\addplot coordinates { (11.1699,10.0459) (11.7549,10.0668) (12.3399,10.0741) (12.9248,10.0922) (13.5098,10.1304) (14.0947,10.129) (14.6797,10.1961) (15.2647,10.2963) (15.8496,10.8396) (16.4346,10.8867) (17.0195,10.8776) (17.6045,11.0676) (18.1895,10.8158) };
\addlegendentry{\iClearyP};
\addplot coordinates { (11.1699,9.39117) (11.7549,9.51714) (12.3399,9.44232) (12.9248,9.62028) (13.5098,9.5027) (14.0947,9.4869) (14.6797,9.55804) (15.2647,9.48429) (15.8496,9.58465) (16.4346,9.6069) (17.0195,9.50396) (17.6045,9.65732) (18.1895,9.54934) };
\addlegendentry{\iClearyS};
\addplot coordinates { (11.1699,9.43136) (11.7549,9.11868) (12.3399,9.61783) (12.9248,9.17084) (13.5098,9.71236) (14.0947,9.30089) (14.6797,9.13554) (15.2647,9.41285) (15.8496,9.27022) (16.4346,9.64145) (17.0195,9.34037) (17.6045,10.0119) (18.1895,9.47863) };
\addlegendentry{\iGoogle};
\addplot coordinates { (11.1699,9.71539) (11.7549,9.72951) (12.3399,9.75272) (12.9248,9.77138) (13.5098,9.80913) (14.0947,9.80295) (14.6797,9.88427) (15.2647,10.0408) (15.8496,10.5384) (16.4346,10.6513) (17.0195,10.6207) (17.6045,10.8275) (18.1895,10.492) };
\addlegendentry{\iLayeredP};
\addplot coordinates { (11.1699,9.01551) (11.7549,9.19517) (12.3399,9.06725) (12.9248,9.25235) (13.5098,9.2031) (14.0947,9.20661) (14.6797,9.28409) (15.2647,9.26835) (15.8496,9.35234) (16.4346,9.28517) (17.0195,9.27022) (17.6045,9.32486) (18.1895,9.29462) };
\addlegendentry{\iLayeredS};
\addplot coordinates { (11.1699,10.6108) (11.7549,11.0014) (12.3399,10.706) (12.9248,10.4754) (13.5098,10.8323) (14.0947,10.5718) (14.6797,10.9491) (15.2647,10.692) (15.8496,10.452) (16.4346,10.9755) (17.0195,10.6015) (17.6045,10.9966) (18.1895,10.7435) };
\addlegendentry{\iPlainD};
\addplot coordinates { (11.1699,6.30256) (11.7549,6.43962) (12.3399,6.27302) (12.9248,6.47789) (13.5098,6.3669) (14.0947,6.25487) (14.6797,6.46434) (15.2647,6.26178) (15.8496,6.4324) (16.4346,6.33151) (17.0195,6.35344) (17.6045,6.27426) (18.1895,6.24539) };
\addlegendentry{\iRigtorp};
\addplot coordinates { (11.1699,7.49905) (11.7549,7.74371) (12.3399,7.54715) (12.9248,7.74236) (13.5098,7.60164) (14.0947,7.58721) (14.6797,7.756) (15.2647,7.64818) (15.8496,7.82972) (16.4346,7.67172) (17.0195,7.6783) (17.6045,7.84528) (18.1895,7.74551) };
\addlegendentry{\iSpp};
\addplot coordinates { (11.1699,8.47438) (11.7549,8.40131) (12.3399,8.44653) (12.9248,8.3936) (13.5098,8.42445) (14.0947,8.43532) (14.6797,8.46448) (15.2647,8.45984) (15.8496,8.42277) (16.4346,8.38183) (17.0195,8.43838) (17.6045,8.38529) (18.1895,8.4363) };
\addlegendentry{\iStd};
\addplot coordinates { (11.1699,7.64506) (11.7549,7.75911) (12.3399,7.71172) (12.9248,7.8157) (13.5098,7.81741) (14.0947,7.75689) (14.6797,7.71859) (15.2647,7.83983) (15.8496,7.91886) (16.4346,7.94701) (17.0195,7.90227) (17.6045,8.6575) (18.1895,7.97423) };
\addlegendentry{\iTsl};

\end{axis}
\end{tikzpicture}
\begin{tikzpicture}
\begin{axis}[
title={Deletion: Time},
xlabel={number of elements [$\lg$]},
ylabel={avg.\ time per element [$\mu$\text{s}/\#]},
appendLegend,
legend to name={legEraseTime},
legend columns=3
]
\addplot coordinates { (11.1699,0.411921) (11.7549,0.434326) (12.3399,0.418243) (12.9248,0.406755) (13.5098,0.426477) (14.0947,0.414445) (14.6797,0.441533) (15.2647,0.428448) (15.8496,0.42304) (16.4346,0.446274) (17.0195,0.436581) (17.6045,0.460837) (18.1895,0.4514) };
\addlegendentry{\iAvxD};
\addplot coordinates { (11.1699,1.24257) (11.7549,1.58087) (12.3399,1.36444) (12.9248,1.15427) (13.5098,1.45274) (14.0947,1.24351) (14.6797,1.58448) (15.2647,1.25264) (15.8496,1.14234) (16.4346,1.42907) (17.0195,1.21135) (17.6045,1.52429) (18.1895,1.30641) };
\addlegendentry{\iChmap};
\addplot coordinates { (11.1699,0.936081) (11.7549,1.1267) (12.3399,0.985565) (12.9248,0.885472) (13.5098,1.03939) (14.0947,0.917141) (14.6797,1.10097) (15.2647,0.965569) (15.8496,0.870221) (16.4346,1.02328) (17.0195,0.969825) (17.6045,1.09037) (18.1895,0.968806) };
\addlegendentry{\iChtI};
\addplot coordinates { (11.1699,0.874727) (11.7549,0.861814) (12.3399,0.865812) (12.9248,0.865616) (13.5098,0.868656) (14.0947,0.865404) (14.6797,0.870207) (15.2647,0.877625) (15.8496,0.885388) (16.4346,0.9204) (17.0195,0.885578) (17.6045,0.898509) (18.1895,0.899168) };
\addlegendentry{\iGoogle};
\addplot coordinates { (11.1699,0.383955) (11.7549,0.406742) (12.3399,0.389134) (12.9248,0.376424) (13.5098,0.396496) (14.0947,0.384676) (14.6797,0.408219) (15.2647,0.395938) (15.8496,0.389587) (16.4346,0.412391) (17.0195,0.410228) (17.6045,0.42907) (18.1895,0.426639) };
\addlegendentry{\iPlainD};
\addplot coordinates { (11.1699,0.257856) (11.7549,0.260127) (12.3399,0.269502) (12.9248,0.278764) (13.5098,0.316615) (14.0947,0.329388) (14.6797,0.336027) (15.2647,0.345473) (15.8496,0.346168) (16.4346,0.355543) (17.0195,0.810513) (17.6045,0.896351) (18.1895,1.3387) };
\addlegendentry{\iRigtorp};
\addplot coordinates { (11.1699,0.998828) (11.7549,1.29076) (12.3399,1.16106) (12.9248,1.41826) (13.5098,1.365) (14.0947,1.18825) (14.6797,1.47513) (15.2647,1.29458) (15.8496,1.59574) (16.4346,1.44595) (17.0195,1.31804) (17.6045,1.6997) (18.1895,1.50019) };
\addlegendentry{\iSpp};
\addplot coordinates { (11.1699,1.02943) (11.7549,1.02809) (12.3399,1.03) (12.9248,1.02862) (13.5098,1.0368) (14.0947,1.04538) (14.6797,1.079) (15.2647,1.11594) (15.8496,1.12319) (16.4346,1.12711) (17.0195,1.14323) (17.6045,1.15268) (18.1895,1.24166) };
\addlegendentry{\iStd};
\addplot coordinates { (11.1699,0.926185) (11.7549,1.18848) (12.3399,1.02765) (12.9248,1.45185) (13.5098,1.23965) (14.0947,0.999228) (14.6797,1.26745) (15.2647,1.11735) (15.8496,1.37063) (16.4346,1.21957) (17.0195,1.06423) (17.6045,1.38195) (18.1895,1.2719) };
\addlegendentry{\iTsl};

\end{axis}
\end{tikzpicture}

\begin{minipage}{0.5\linewidth}
\ref{legMissTime}
\end{minipage}
\begin{minipage}{0.5\linewidth}
\hfill
\ref{legEraseTime}
\end{minipage}

\caption{\emph{Left}: Time for looking up $2^8$ random keys that are not present in the hash tables.
\emph{Right}: Time for erasing $2^8$ random keys that are present in the hash tables.
In both figures, the number of elements (x-axis) is the number of elements a hash table contains.}
\label{figEraseTime}
\end{figure*}

\myblock{Unsuccessful Searches}
The search of not stored elements is far more time consuming with our separate chaining hash tables, as can be seen in the left of \cref{figEraseTime}.
When restricted to separate chaining, best bets can be made with \iAvx{},
as it is the fastest for scanning large buckets.
It is on a par with the Bonsai tables, but no match for the sparse hash tables.

\myblock{Removing Elements}
We evaluated the speed for removing arbitrary elements from the hash tables, and present the results in the right of \cref{figEraseTime}.
We used the hash tables created during the construction benchmark (\cref{figRandomInsertTime}).
Interestingly, \iAvx{} becomes faster than \iRigtorp{} in the last instance.
The other implementations are on a par with the non-compact sparse contestants.
We could not evaluate the Bonsai tables, as there is currently no implementation available for removing elements.

\begin{figure*}

\pgfplotsset{squeezedPlot}

\begin{tikzpicture}
\begin{axis}[
title={\texttt{UDB2}: Space},
xlabel={number of elements [$\lg$]},
ylabel={memory [MiB]}
]
\addplot coordinates { (23.2535,14) (24.1015,26) (24.632,37) (25.019,48) (25.3239,59) (25.5754,70) };
\addlegendentry{\iChmapI};
\addplot coordinates { (23.2535,14) (24.1015,26) (24.632,37) (25.019,48) (25.3239,60) (25.5754,71) };
\addlegendentry{\iChtI};
\addplot coordinates { (23.2535,20) (24.1015,35) (24.632,41) (25.019,74) (25.3239,82) (25.5754,120) };
\addlegendentry{\iClearyS};
\addplot coordinates { (23.2535,20) (24.1015,36) (24.632,51) (25.019,69) (25.3239,84) (25.5754,98) };
\addlegendentry{\iGoogle};
\addplot coordinates { (23.2535,22) (24.1015,39) (24.632,45) (25.019,82) (25.3239,90) (25.5754,143) };
\addlegendentry{\iLayeredS};
\addplot coordinates { (23.2535,18) (24.1015,34) (24.632,49) (25.019,64) (25.3239,79) (25.5754,94) };
\addlegendentry{\iPlainI};
\addplot coordinates { (23.2535,26) (24.1015,53) (24.632,62) (25.019,77) (25.3239,108) (25.5754,122) };
\addlegendentry{\iSpp};
\addplot coordinates { (23.2535,24) (24.1015,47) (24.632,59) (25.019,74) (25.3239,100) (25.5754,115) };
\addlegendentry{\iTsl};

\legend{}
\end{axis}
\end{tikzpicture}
\begin{tikzpicture}
\begin{axis}[
title={\texttt{UDB2}: Time/Space },
xlabel={avg.\ time per element [s/\#]},
ylabel={avg.\ memory per element [byte/\#]},
appendLegend,
legend to name={legUDBTime},
legend columns=2
]
\addplot coordinates { (0.5075,1.54955) (0.533885,1.51557) (0.539778,1.52067) (0.544167,1.48727) (0.54926,1.48506) (0.560441,1.49048) };
\addlegendentry{\iChmapI};
\addplot coordinates { (0.216,1.56266) (0.249611,1.53523) (0.260038,1.52565) (0.274029,1.5059) (0.281286,1.49975) (0.295,1.49555) };
\addlegendentry{\iChtI};
\addplot coordinates { (0.2798,2.10928) (0.3195,1.68676) (0.345333,2.06384) (0.406167,2.04989) (0.428824,2.308176) (0.5383,2.5198) };
\addlegendentry{\iClearyS};
\addplot coordinates { (0.1335,2.10358) (0.154389,2.11901) (0.163269,2.07148) (0.171882,2.12862) (0.173143,2.09742) (0.17872,2.076172) };
\addlegendentry{\iGoogle};
\addplot coordinates { (0.2781,2.31939) (0.313192,1.84866) (0.344111,2.29691) (0.396286,2.25006) (0.425029,2.55502) (0.52778,3.01604) };
\addlegendentry{\iLayeredS};
\addplot coordinates { (0.1952,1.99162) (0.229944,1.99461) (0.268731,2.00679) (0.270294,1.9969) (0.273167,1.99079) (0.27744,1.98661) };
\addlegendentry{\iPlainI};
\addplot coordinates { (0.1195,2.8216) (0.145222,3.13454) (0.156154,2.53918) (0.157618,2.39657) (0.18112,2.56597) (0.182024,2.70732) };
\addlegendentry{\iSpp};
\addplot coordinates { (0.0909,2.54092) (0.113833,2.73817) (0.121231,2.40749) (0.1255,2.3035) (0.15231,2.51334) (0.1532,2.42607) };
\addlegendentry{\iTsl};

\end{axis}
\end{tikzpicture}

\begin{minipage}{0.5\textwidth}
\label{figUDBTime}
\caption{%
\parbox{0.9\linewidth}{%
Time for processing hashed keys from a random generator provided in the \texttt{udb2} benchmark. Keys and values are 32 bit integers.
}
}
\end{minipage}
\begin{minipage}{0.5\textwidth}
\hfill
\ref{legUDBTime}
\end{minipage}

\end{figure*}
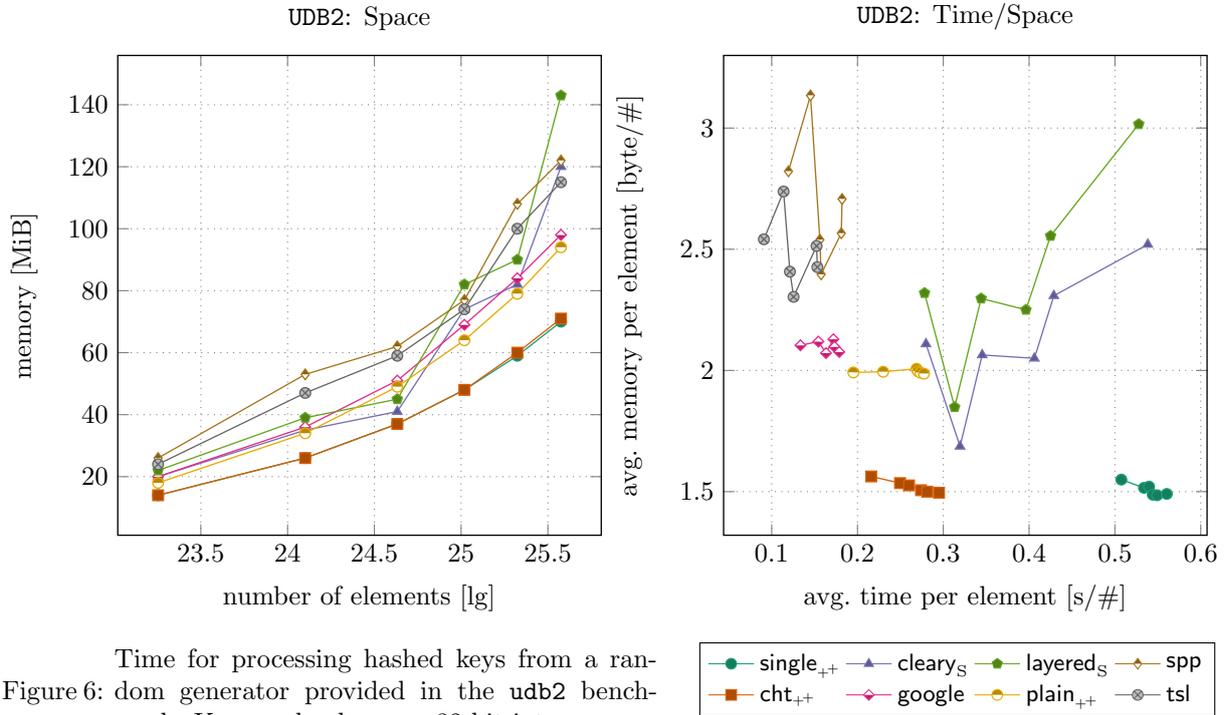

\myblock{Distinct Keys}
We inserted our hash tables into the \texttt{udb2} benchmark\footnote{\url{https://github.com/attractivechaos/udb2}},
where the task is to compute the frequencies of all 32-bit keys of a multiset, 
in which roughly $25\%$ of all keys are distinct.
For that, the hash tables store each of these keys along with a 32-bit value counting its processed occurrences.
Our results are shown in \cref{figUDBTime}.
We expect from a succinct representation to use space about $2$ bytes per key, as about $25\%$ of all keys are distinct, and each (key,value)-pair takes $8$ bytes.
The evaluation shows that, if time is not of importance, the memory footprint can be considerably improved with our proposed hash table layout.

\subsection{Conclusion}
On the upside, the evaluation reveals that our proposed hash tables can be constructed at least as fast as all other compact sparse hash tables (cf.~\cref{figRandomInsertTime}).
Our hash tables use less space than any non-sparse compact hash table (cf.~\cref{figRandomInsertConstruction}).
Especially fast are deletions (cf.~\cref{figEraseTime}), outpacing even some speed-optimized hash tables on large instances.
Combining \iAvxD{}  with compact hashing can lead to a fast and memory-efficient hash table  if there are good lower bounds on the number of elements that need to be stored (cf.~\cref{figRandomInsertTime} for the time and \iPlainD{} in \cref{figMemoryReserve} for the space).

On the downside,
lookups, especially when searching for a non-present key, are even slower than most of the sparse Bonsai tables, as \bucketMax{} is much larger than the number of maximal collisions encountered during an insertion of an element in one of the Bonsai tables.
That is because their default maximum load factor of $\alpha := 0.5$ gives $c_\alpha \le 3$ collisions on average for an insertion operation (cf.~\cref{eqCollision}).

In total, the major advantage of our proposed hash table layout is its low memory footprint.
Its construction speed matches with other memory-efficient hash table representations.
However, if the focus of an application is on querying rather than on dynamic aspects as insertion or deletion,
Cuckoo hash tables or perfect hashing provide a much better solution.

\subsection{Future Work}
We think that a bucketized compact Cuckoo hash table~\cite{ross07simdhash} based on our proposed hash table layout can be an even more memory-friendly hash table.
For that, we store a master and a slave separate chaining hash table whose numbers of buckets are independent from each other.
On inserting an element, we first try to insert the element in the master table.
If its respective bucket~$B_\textup{M}$ is already full, we try to insert the element in the slave table.
If its respective bucket~$B_\textup{S}$ is also full, we take a random element of both buckets~$B_\textup{M}$ and~$B_\textup{S}$,
exchange it with the element we want to insert, and start a random walk.
By doing so, the distribution of the load factors of all buckets should become more uniform such that a resizing of the hash tables can be delayed at the expense of more comparisons.
Both hash tables can be made compact, as each bucket is dedicated to exactly one injective transform (corresponding to its respective hash table).

In the experiments, the measured memory is the number of allocated bytes.
The resident set sizes of our hash tables differ largely to this measured memory, as we allocate many tiny fragments of space.
A dedicated memory manager can reduce this space overhead, but also reduce the memory requirement of the bucket pointers by allocating a large subsequent array, in which memory can be addressed with pointers of 32-bit width or less.
For future evaluations, we also want to vary the maximum load factors of all hash tables instead of sticking to the default ones.

For searching data in an array,
the more recent SIMD instruction set AVX2 provides a major performance boost unlike older instruction sets like SSE,
as benchmarks for comparing strings\footnote{\url{https://github.com/koeppl/packed_string}} demonstrate a speed boost of more than 50\% for long string instances.
We wonder whether we can experience an even steeper acceleration when working with the AVX256 instruction set.

In our implementation of \iCht{}, we extract the quotients from its bit-compact byte array~$B$ \emph{sequentially} during the search of a quotient~$q$.
We could accelerate this search by packing~$q$ $\gauss{64/k}$ times in one 64-bit integer~$p$, 
where $k$ is the quotient bit width, 
and compare the same number of quotients in~$B$ with $B[i..i+63] \otimes p$ for $i = c k\gauss{64/k}$ with an integer~$c$,
where we interpret $B$ as a bit vector.
Using shift and bitwise AND operations, we can compute a bit vector~$C$ such that $C[j] = 1 \Leftrightarrow q = B[i+(j-1)k..i+jk-1]$ for $1 \le j \le \gauss{64/k}$, in \Oh{\lg k} time by using bit parallelism.

Finally, we would like to see our hash table in applications where saving space is critical.
For instance, we could devise the Bonsai trie~\cite{darragh93bonsai} or the displacement array of \iLayered{}~\cite{poyias15bonsai}, which are used, for instance, in the LZ78 computation~\cite{arroyuelo17lz78}.

\begin{acknowledgment}
We are thankful to Rajeev Raman for a discussion about future work on compact hash tables at the Dagstuhl seminar~18281,
to Marvin Löbel for the implementations of the Bonsai hash tables,
to Shunsuke Kanda for the implementation of our used injective transform,
and to Manuel Penschuck for running the entropy experiments on a computing cluster.

This work is founded by the JSPS KAKENHI Grant Number JP18F18120.
\end{acknowledgment}

\ifcsname newblock\endcsname
\else
\newcommand{\newblock}{}
\fi

\bibliographystyle{abbrvnat}
\bibliography{literature,references}

\renewcommand\floatpagefraction{\oldfloatpagefraction}
\renewcommand\dblfloatpagefraction{\olddblfloatpagefraction} %

\renewcommand\floatpagefraction{\oldfloatpagefraction}
\renewcommand\dblfloatpagefraction{\olddblfloatpagefraction} %

\appendix

\section{Applications}
We provide two applications of our hash tables on real-world data sets.

\subsection{Keyword Fingerprinting}
An application of hash tables is to store fingerprints of a set of keywords.
We hash each keyword with Austin Appleby's Murmurhash\footnote{\url{https://github.com/aappleby/smhasher}}, which is a well received hash function for generating fingerprints~\cite{fu17murmur,nair18murmur}.
The obtained fingerprint is put into the hash tables.
Non-compact hash tables use the identity as a dummy hash function, while
the compact hash tables use our injective transform.
Such a hash table can be used as a Bloom filter~\cite{bloom70bloomfilter} for discarding strings that are not part of the set of keywords.
\Cref{tableFingerprintInsert} gives the time and space needed for constructing such a Bloom filter.
In \Cref{tableFingerprintQuery} we measure the time it takes to query for all inserted keywords.
We used data sets from~\cite{takagi17packedtrie} and~\cite{dinklage17tudocomp}, 
split each dataset into strings by using either the newline or whitespace as a delimiter, and removed all duplicates.
We can see that our separate chaining hash table variants use far less space then the compact non-sparse hash tables,
while they are smaller or on par with their sparse variants.

\begin{table*}

\begin{tabular}{l*{5}{r}}
\toprule
 & \multicolumn{5}{c}{data sets} \\
\cmidrule(lr){2-6}
 hash table & \texttt{cc} & \texttt{dblp} & \texttt{proteins} & \texttt{urls} & \texttt{wiki} \\
\midrule
\iChmap &  21.1  &  29.0  &  29.3  &  176.0  &  5.0  \\
\iChtD &  25.0  &  34.9  &  35.0  &  218.5  &  6.0  \\
\iChtI &  21.2  &  29.1  &  29.5  &  177.0  &  5.0  \\
\iClearyS &  21.3  &  30.2  &  30.5  &  190.8  &  5.1  \\
\iEliasP &  57.5  &  112.0  &  112.0  &  859.7  &  14.6  \\
\iEliasS &  21.0  &  29.5  &  29.8  &  184.8  &  5.0  \\
\iGoogle &  33.2  &  46.4  &  46.9  &  293.9  &  7.9  \\
\iLayeredP &  59.5  &  116.0  &  116.1  &  892.3  &  15.1  \\
\iLayeredS &  22.3  &  32.2  &  32.5  &  212.4  &  5.3  \\
\iPlainD &  29.0  &  40.8  &  40.9  &  265.1  &  6.9  \\
\iPlainI &  24.6  &  34.0  &  34.4  &  214.6  &  5.7  \\
\iRigtorp &  96.7  &  192.0  &  192.0  &  1536.0  &  24.0  \\
\iStd &  85.8  &  95.9  &  96.7  &  706.0  &  20.5  \\
\iTsl &  35.4  &  52.0  &  52.5  &  339.3  &  8.4  \\
\bottomrule
\end{tabular}
\hfill
\begin{tabular}{l*{5}{r}}
\toprule
 & \multicolumn{5}{c}{data sets} \\
\cmidrule(lr){2-6}
 hash table & \texttt{cc} & \texttt{dblp} & \texttt{proteins} & \texttt{urls} & \texttt{wiki} \\
\midrule
\iChmap &  3.8  &  5.2  &  5.5  &  36.2  &  1.0  \\
\iChtD &  1.5  &  2.0  &  2.3  &  14.4  &  0.4  \\
\iChtI &  1.9  &  2.7  &  3.0  &  21.1  &  0.5  \\
\iClearyS &  1.2  &  2.0  &  2.4  &  16.0  &  0.3  \\
\iEliasP &  5.7  &  8.9  &  9.3  &  63.3  &  1.5  \\
\iEliasS &  6.1  &  9.6  &  10.1  &  70.1  &  1.5  \\
\iGoogle &  1.0  &  1.4  &  1.8  &  12.3  &  0.3  \\
\iLayeredP &  0.6  &  0.9  &  1.2  &  7.1  &  0.2  \\
\iLayeredS &  1.2  &  1.9  &  2.3  &  15.7  &  0.3  \\
\iPlainD &  0.8  &  1.1  &  1.5  &  9.3  &  0.2  \\
\iPlainI &  1.2  &  1.8  &  2.1  &  16.8  &  0.3  \\
\iRigtorp &  0.3  &  0.5  &  0.8  &  3.7  &  0.1  \\
\iStd &  0.9  &  1.2  &  1.5  &  9.5  &  0.3  \\
\iTsl &  0.4  &  0.7  &  1.0  &  6.2  &  0.2  \\
\bottomrule
\end{tabular}

\begin{minipage}{0.5\linewidth}
   \centering{memory [MiB]}
\end{minipage}
\begin{minipage}{0.5\linewidth}
   \centering{time [s]}
\end{minipage}

\caption{Construction of a fingerprint keyword dictionary.}
\label{tableFingerprintInsert}
\end{table*}

\begin{table}
\centerline{%
\begin{tabular}{l*{5}{r}}
\toprule
 & \multicolumn{5}{c}{data sets} \\
\cmidrule(lr){2-6}
 hash table & \texttt{cc} & \texttt{dblp} & \texttt{proteins} & \texttt{urls} & \texttt{wiki} \\
\midrule
\iChmap &  0.4  &  0.8  &  1.1  &  5.7  &  0.2  \\
\iChtD &  0.5  &  1.0  &  1.3  &  7.4  &  0.2  \\
\iChtI &  0.4  &  0.8  &  1.1  &  5.7  &  0.2  \\
\iClearyS &  0.4  &  0.6  &  0.9  &  5.3  &  0.2  \\
\iEliasP &  2.6  &  3.2  &  3.5  &  18.8  &  0.7  \\
\iEliasS &  2.6  &  3.1  &  3.5  &  18.4  &  0.7  \\
\iGoogle &  0.3  &  0.5  &  0.8  &  4.4  &  0.1  \\
\iLayeredP &  0.3  &  0.5  &  0.7  &  3.3  &  0.1  \\
\iLayeredS &  0.4  &  0.6  &  0.9  &  5.2  &  0.2  \\
\iPlainD &  0.5  &  0.9  &  1.2  &  6.9  &  0.1  \\
\iPlainI &  0.5  &  0.9  &  1.3  &  6.9  &  0.1  \\
\iRigtorp &  0.3  &  0.4  &  0.6  &  2.7  &  0.1  \\
\iStd &  0.5  &  0.8  &  1.0  &  5.2  &  0.2  \\
\iTsl &  0.3  &  0.5  &  0.8  &  4.0  &  0.1  \\
\bottomrule
\end{tabular}
}%
\caption{Query time in seconds for a fingerprint keyword dictionary.}
\label{tableFingerprintQuery}
\end{table}

\subsection{Computing the Empirical Entropy}
\newcommand*{\textT}  {\ensuremath{T}}
\newcommand*{\textS}  {\ensuremath{S}}

Given a text~$T$ of length~$n$ whose characters are drawn from a finite alphabet~$\Sigma := \menge{c_1, \ldots, c_\sigma}$,
the empirical entropy~$H_k$ of order~$k$ for an integer $k \ge 0$ is defined as
$H_0(\textT) := (1/n)\sum_{j=1}^{\sigma} n_j \lg(n/n_j)$ for $n_j := \abs{\menge{i : \textT[i] = c_j}}$,
and
$H_k(\textT) := (1/n) \sum_{\textS \in \Sigma^k} \abs{\textT_{\textS}} H_0(\textT_{\textS})$,
where $\textT_{\textS}$ is the concatenation of each character in $\textT$ that directly follows an occurrence of the substring~$\textS \in \Sigma^k$ in $\textT$.
We can compute $H_0(T)$ with an array using $\sigma \upgauss{\lg n}$ bits of space storing the frequency of each character.
For larger orders, we count all $k$-mers, i.e., substrings of length~$k$, in $T$ and iterate over the frequencies of all $k$-mers of $T$ to compute~$H_k(T)$.
Using an array with $\sigma^k \upgauss{\lg n}$ bits, this approach can become obstructive for large alphabet sizes.
For small alphabets like in DNA sequences, highly optimized $k$-mer counters can compute the entropy up to order~55~\cite{manekar18kmer}.

Here, we present an approach that stores the frequencies of the $k$-mers with our separate chaining hash table.
Our approach is similar to Jellyfish~\cite{marcais11jellyfish}, but more naive as we do not apply Bloom filters or concurrency for speed-up.
Instead, our target is to compute the entropy for byte alphabets, orders $k \le 7$, but massive data sets.
For that task, we use \iChtI{} and start with byte values representing the frequencies.
Whenever the current representation of the frequencies becomes too small, we increment the number of bytes of the frequency representation by one.
By doing so, we reach up to 3 bytes per stored frequency in our experiments.
As the experimental setup is also of independent interest for computing the empirical entropy of massive data sets,
we made it freely available at \url{https://github.com/koeppl/compression_indicators}.

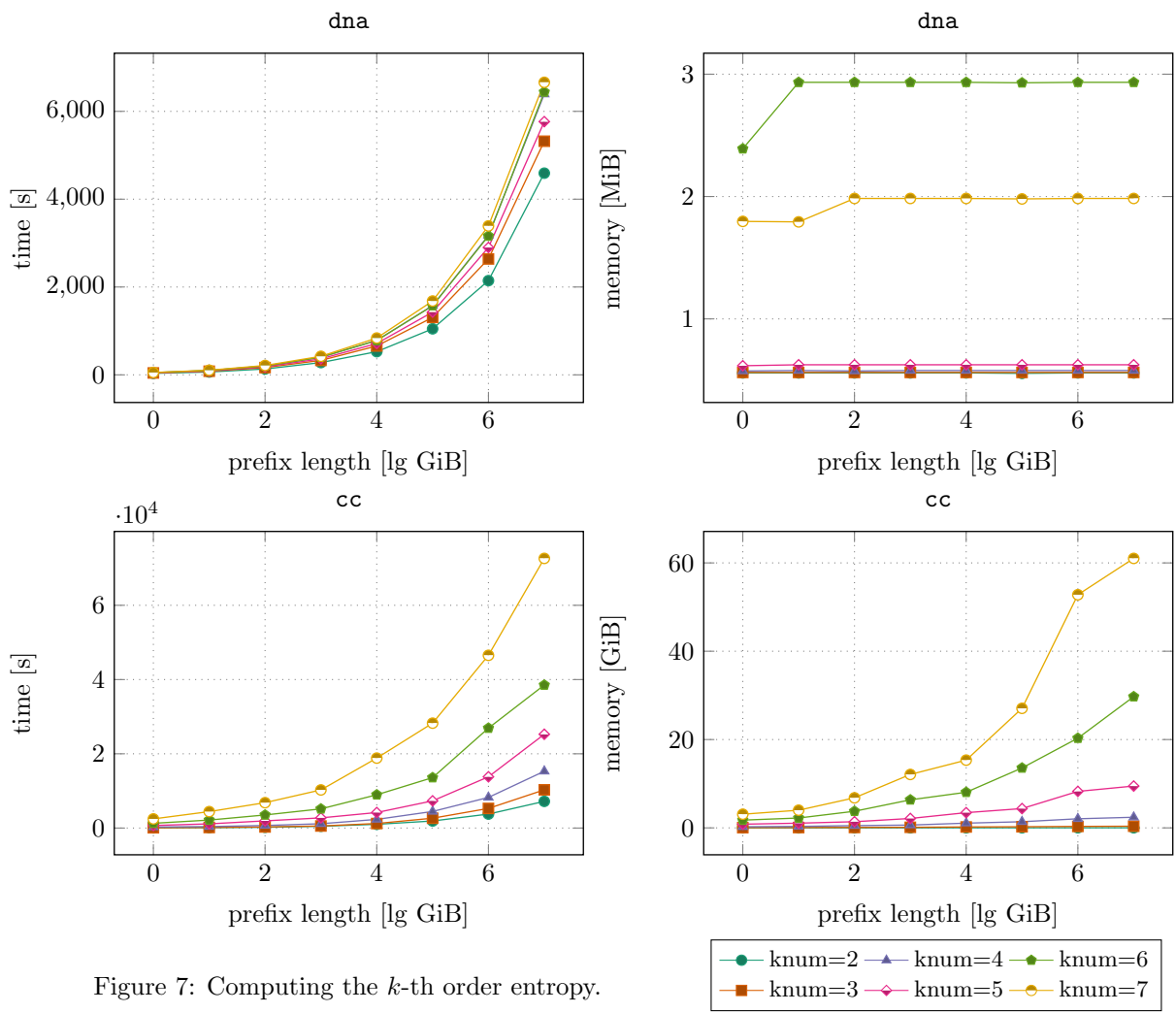
\begin{figure*}

\pgfplotsset{squeezedPlot, height=0.4\linewidth}

\begin{tikzpicture}
\begin{axis}[title={\texttt{dna}},
xlabel={prefix length [$\lg$ GiB]},
ylabel={time [s]},
appendLegend,
legend to name={legEntropy},
legend columns=2
]

\addplot coordinates { (0.0,33) (1.0,65) (2.0,133) (3.0,279) (4.0,528) (5.0,1048) (6.0,2145) (7.0,4592) };
\addlegendentry{knum=2};
\addplot coordinates { (0.0,43) (1.0,82) (2.0,165) (3.0,329) (4.0,657) (5.0,1307) (6.0,2639) (7.0,5320) };
\addlegendentry{knum=3};
\addplot coordinates { (0.0,50) (1.0,100) (2.0,200) (3.0,399) (4.0,784) (5.0,1573) (6.0,3176) (7.0,6393) };
\addlegendentry{knum=4};
\addplot coordinates { (0.0,45) (1.0,89) (2.0,182) (3.0,364) (4.0,713) (5.0,1429) (6.0,2900) (7.0,5766) };
\addlegendentry{knum=5};
\addplot coordinates { (0.0,50) (1.0,99) (2.0,200) (3.0,402) (4.0,783) (5.0,1575) (6.0,3161) (7.0,6441) };
\addlegendentry{knum=6};
\addplot coordinates { (0.0,53) (1.0,105) (2.0,213) (3.0,422) (4.0,838) (5.0,1681) (6.0,3391) (7.0,6661) };
\addlegendentry{knum=7};

\end{axis}
\end{tikzpicture}
\begin{tikzpicture}
\begin{axis}[
title={\texttt{dna}},
xlabel={prefix length [$\lg$ GiB]},
ylabel={memory [MiB]}
]

\addplot coordinates { (0.0,0.558594) (1.0,0.558594) (2.0,0.558594) (3.0,0.558594) (4.0,0.558594) (5.0,0.554688) (6.0,0.558594) (7.0,0.558594) };
\addlegendentry{knum=2};
\addplot coordinates { (0.0,0.5625) (1.0,0.5625) (2.0,0.5625) (3.0,0.5625) (4.0,0.5625) (5.0,0.5625) (6.0,0.5625) (7.0,0.5625) };
\addlegendentry{knum=3};
\addplot coordinates { (0.0,0.574219) (1.0,0.578125) (2.0,0.574219) (3.0,0.578125) (4.0,0.578125) (5.0,0.578125) (6.0,0.578125) (7.0,0.578125) };
\addlegendentry{knum=4};
\addplot coordinates { (0.0,0.617188) (1.0,0.625) (2.0,0.625) (3.0,0.625) (4.0,0.625) (5.0,0.625) (6.0,0.625) (7.0,0.625) };
\addlegendentry{knum=5};
\addplot coordinates { (0.0,2.390625) (1.0,2.93359) (2.0,2.93359) (3.0,2.93359) (4.0,2.93359) (5.0,2.92969) (6.0,2.93359) (7.0,2.93359) };
\addlegendentry{knum=6};
\addplot coordinates { (0.0,1.796875) (1.0,1.79297) (2.0,1.984375) (3.0,1.984375) (4.0,1.984375) (5.0,1.98047) (6.0,1.984375) (7.0,1.984375) };
\addlegendentry{knum=7};

\legend{}; %
\end{axis}
\end{tikzpicture}

\begin{tikzpicture}
\begin{axis}[
width=0.5\linewidth,
height=6cm,
title={\texttt{cc}},
xlabel={prefix length [$\lg$ GiB]},
ylabel={time [s]}
]

\addplot coordinates { (0.0,53) (1.0,102) (2.0,195) (3.0,476) (4.0,1005) (5.0,1897) (6.0,3761) (7.0,7185) };
\addlegendentry{knum=2};
\addplot coordinates { (0.0,79) (1.0,152) (2.0,291) (3.0,551) (4.0,1216) (5.0,2646) (6.0,5285) (7.0,10289) };
\addlegendentry{knum=3};
\addplot coordinates { (0.0,278) (1.0,392) (2.0,684) (3.0,1155) (4.0,2326) (5.0,4466) (6.0,8247) (7.0,15297) };
\addlegendentry{knum=4};
\addplot coordinates { (0.0,684) (1.0,1125) (2.0,1928) (3.0,2735) (4.0,4178) (5.0,7319) (6.0,13802) (7.0,25257) };
\addlegendentry{knum=5};
\addplot coordinates { (0.0,1275) (1.0,2148) (2.0,3523) (3.0,5160) (4.0,8956) (5.0,13580) (6.0,26914) (7.0,38520) };
\addlegendentry{knum=6};
\addplot coordinates { (0.0,2459) (1.0,4425) (2.0,6834) (3.0,10267) (4.0,18839) (5.0,28252) (6.0,46527) (7.0,72636) };
\addlegendentry{knum=7};

\legend{}; %
\end{axis}
\end{tikzpicture}
\begin{tikzpicture}
\begin{axis}[
width=0.5\linewidth,
height=6cm,
title={\texttt{cc}},
xlabel={prefix length [$\lg$ GiB]},
ylabel={memory [GiB]}
]

\addplot coordinates { (0.0,0.00542831) (1.0,0.00712585) (2.0,0.0080986) (3.0,0.0109482) (4.0,0.0127525) (5.0,0.0150414) (6.0,0.0168495) (7.0,0.0177994) };
\addlegendentry{knum=2};
\addplot coordinates { (0.0,0.0507812) (1.0,0.073307) (2.0,0.0910072) (3.0,0.128498) (4.0,0.203087) (5.0,0.251415) (6.0,0.328045) (7.0,0.373676) };
\addlegendentry{knum=3};
\addplot coordinates { (0.0,0.247303) (1.0,0.327999) (2.0,0.506977) (3.0,0.636135) (4.0,1.06596) (5.0,1.3892) (6.0,2.05634) (7.0,2.40339) };
\addlegendentry{knum=4};
\addplot coordinates { (0.0,0.828838) (1.0,1.06199) (2.0,1.39116) (3.0,2.12627) (4.0,3.44331) (5.0,4.36792) (6.0,8.2579) (7.0,9.46931) };
\addlegendentry{knum=5};
\addplot coordinates { (0.0,1.75264) (1.0,2.23042) (2.0,3.76005) (3.0,6.34815) (4.0,8.06007) (5.0,13.5662) (6.0,20.3078) (7.0,29.7105) };
\addlegendentry{knum=6};
\addplot coordinates { (0.0,3.09052) (1.0,4.00109) (2.0,6.80362) (3.0,12.0944) (4.0,15.3386) (5.0,27.0671) (6.0,52.7884) (7.0,61.0206) };
\addlegendentry{knum=7};

\legend{}; %
\end{axis}
\end{tikzpicture}

\begin{minipage}{0.6\textwidth}
\caption{Computing the $k$-th order entropy. }
\label{figEntropy}
\end{minipage}
\begin{minipage}{0.4\textwidth}
\ref{legEntropy}
\end{minipage}

\end{figure*}

For the experiments, we took two datasets, \texttt{cc} and \texttt{dna}, each of $128$ GiB.
The former data set has an alphabet size of 242, and consists of a web page crawl provided by the commoncrawl organization\footnote{\url{http://commoncrawl.org/}}.
The latter data set is a collection of DNA sequences extracted from FASTA files with an alphabet size of 4.
We computed the entropies of each prefix of length $2^n (1024)^3$, for $1 \le n \le 7$, for the data sets \texttt{cc} and \texttt{dna} in \cref{tableEntropyCC} and \cref{tableEntropyDNA}, respectively.
We summarize the needed time and space for these computations in \cref{figEntropy}.
These experiments ran on a computing cluster equipped with Intel Xeon Gold 6148 CPUs clocked at 2.40GHz with 192 GiB of RAM running Red Hat Linux~4.8.5-36.
The measured memory is the maximum used resident set size.

We can see that the amount of needed memory becomes saturated after processing the first 2 GiB of \texttt{dna}, where we use less than $3$ MiB of RAM in total for all orders of~$k$.
That is not surprising, as there can be only $4^7$ different $k$-mers of length $7$.
For \texttt{cc}, it is more relevant to have a memory-efficient implementation, as there can be $242^7 \approx 5 \cdot 10^{10}$ $7$-mers.
We conclude by the strictly monotonic increase of the occupied memory that new $k$-mers for $k \ge 4$ are found in \texttt{cc} even after surpassing the $64$ GiB prefix.

\begin{table}
   \setlength{\tabcolsep}{2pt}
\begin{tabular}{l*{6}{r}}
\toprule
prefix & \multicolumn{6}{c}{order $k$} \\
\cmidrule(lr){2-7}
length & {2} & {3} & {4} & {5} & {6} & {7} \\
\midrule
1 &  3.47259  &  2.90481  &  2.35796  &  1.90477  &  1.49418  &  1.18580  \\
2 &  3.48268  &  2.92203  &  2.38605  &  1.94619  &  1.54430  &  1.23745  \\
4 &  3.48717  &  2.93171  &  2.40566  &  1.97859  &  1.58677  &  1.28333  \\
8 &  3.48762  &  2.93742  &  2.41886  &  2.00233  &  1.62009  &  1.32094  \\
16 &  3.48920  &  2.94113  &  2.42738  &  2.01886  &  1.64558  &  1.35130  \\
32 &  3.49006  &  2.94411  &  2.43471  &  2.03284  &  1.66737  &  1.37798  \\
64 &  3.49100  &  2.94669  &  2.44088  &  2.04409  &  1.68482  &  1.40001  \\
128 &  3.49055  &  2.94684  &  2.44231  &  2.04753  &  1.69087  &  1.40839  \\
\bottomrule
\end{tabular}
\caption{Empirical entropy of the data set \texttt{cc}. Prefix length is in GiB.}
\label{tableEntropyCC}
\end{table}

\begin{table}
   \setlength{\tabcolsep}{2pt}
\begin{tabular}{l*{6}{r}}
\toprule
prefix & \multicolumn{6}{c}{order $k$} \\
\cmidrule(lr){2-7}
length & {2} & {3} & {4} & {5} & {6} & {7} \\
\midrule
1 &  1.94051  &  1.86247  &  1.77680  &  1.69265  &  1.61940  &  1.56313  \\
2 &  1.91210  &  1.87496  &  1.83286  &  1.78250  &  1.73037  &  1.68750  \\
4 &  1.92923  &  1.91052  &  1.88797  &  1.85679  &  1.82093  &  1.78944  \\
8 &  1.93363  &  1.92250  &  1.90831  &  1.88764  &  1.86061  &  1.83383  \\
16 &  1.93166  &  1.92232  &  1.91167  &  1.89491  &  1.87101  &  1.84585  \\
32 &  1.93201  &  1.92421  &  1.91507  &  1.90190  &  1.88270  &  1.86160  \\
64 &  1.93145  &  1.92424  &  1.91588  &  1.90445  &  1.88763  &  1.86889  \\
128 &  1.93873  &  1.93273  &  1.92486  &  1.91341  &  1.89601  &  1.87634  \\
\bottomrule
\end{tabular}
\caption{Empirical entropy of the data set \texttt{dna}. Prefix length is in GiB.}
\label{tableEntropyDNA}
\end{table}

\end{document}